# Analytical models for the design of photophoretically levitating macroscopic sensors in the stratosphere

Benjamin C. Schafer[1,*], Jong-hyoung Kim[1], Joost J. Vlassak[1], David W. Keith[1]

[1]John A. Paulson School of Engineering and Applied Sciences, Harvard University, Cambridge, MA 02138

*Email: schafer@g.harvard.edu

**Abstract**

Photophoretic forces could levitate thin 10 centimeter-scale structures in Earth's stratosphere indefinitely. We develop analytical models of the thermal transpiration force on a bilayer sandwich structure in the stratosphere. Lofting is maximized when the layers are separated by an air gap equal to the mean free path (MFP), when about half of the layers' surfaces consist of holes with radii < MFP, and when the top layer is solar-transmissive and infrared-emissive while the bottom layer is solar-absorptive and infrared-transmissive. We use the models to design a 10 cm diameter device with sufficient strength to withstand forces that might be encountered in transport, deployment, and flight. The device has a payload of about 285 mg at an altitude of 25 km; enough to support bidirectional radio communication at over 10 Mb/s and limited navigation. Such devices could be useful for atmospheric science or telecommunications on Earth and Mars. Structures a few times larger might have payloads of a few grams.

**Main text**

**Introduction**

Differences in temperature or accommodation coefficient can induce gas flows when the mean free path (MFP) of the gas is large compared to the distance over which the temperature or accommodation coefficients vary. Such flows have been investigated since Crookes demonstrated his radiometer in 1873. Practical applications of this phenomenon are growing; they now include holographic displays [1], gas pumps without moving parts [2], and photophoretic particle traps [3]. Keith [4] (2010) suggested that microscale, engineered particles could self-levitate in the stratosphere using $\Delta\alpha$ photophoresis, a force caused by an asymmetry in the accommodation coefficients of the particles' top and bottom surfaces. Azadi et al. [5] (2020) experimentally demonstrated levitation of a 6 mm illuminated disk using the phenomenon described by Keith. But a bilayer isothermal structure using this phenomena can only levitate if the horizontal size is small compared to the MFP, limiting practical structures in the mid-stratosphere to less than 1 mm. Cortes et al. [6] (2020) demonstrated levitation of structures roughly $10^3$ times larger than those proposed by Keith and realized by Azadi et al. Cortes et al. used fabricated hollow alumina sandwich structures of order 1 cm wide, order 100 µm thick, and with area density 1 g/m², termed "nanocardboard," that develop internal temperature gradients when illuminated. These gradients generate upward thrust via thermal transpiration, or thermal "creep" flow, through vertical channels spaced periodically throughout the structures. Thermal transpiration can generate much larger total upward forces than $\Delta\alpha$ photophoresis at a given pressure. When large payload capacity is desired, nanocardboard-like structures are more appropriate for atmospheric flight than the microscale structures of Keith and Azadi et al.

Here we develop simple physical models to predict the levitating force on a nanocardboard-like structure in the stratosphere. We then use these models to provide a preliminary design of a practical device large enough to communicate with the ground. Our design is constrained to structures that could be fabricated using current nano-fabrication tools. In Kim et al. [7], we discuss the fabrication, testing, and structural analysis of the nanocardboard-like structures used in our practical device. Such a device could expand capabilities for in-situ atmospheric sensing, remote measurements, and telecommunications to include microscale technologies that might complement existing aircraft and balloons.

We first present our 1-dimensional analytical models, which calculate the lofting force on a structure with specified radiative, thermal, and design parameters at a given altitude with typical equatorial radiative forcing. The middle stratosphere is a good environment for levitation due to its minimal turbulence and slow settling velocity (of order 1 cm/s) for centimeter-scale thin plates weighing hundreds of mg. Our proposed disk-shaped device has a 10 cm diameter and could levitate at an altitude of 25 km while carrying a 285 mg payload. We then provide rough estimates that suggest that existing technologies could be adapted to enable attitude adjustment, motion control, deployment, and communication with the ground. Lastly, we explore sensing and telecommunication applications on Earth and lofting on Mars.

**Background.** Some background on photophoresis is useful for understanding our models. Radiative heat transfer induces thermal gradients within a structure and between the structure and its surroundings. The thermal gradients induce gas flows, and these flows modify the thermal gradients. The resulting photophoretic forces on a structure may be categorized into three distinct phenomena: (1) a difference in the thermal accommodation coefficient $\alpha$ over the surface of an isothermal structure of different temperature than the ambient gas causes a net momentum transfer to the structure via gas collisions ($\Delta\alpha$ photophoresis, Fig. 1a), (2) a difference in the temperature over the surface of the structure also causes a net momentum transfer to the structure via gas collisions ($\Delta T$ photophoresis, Fig. 1b), and (3) a thermal gradient between two regions of gas creates a flow from the cold region to the hot region, which transfers momentum to a nearby structure (thermal transpiration, Fig. 1c) [8,9].

Photophoretic forces are only significant when the gas flows that impart them are in the free-molecular (FM) regime, defined roughly as where the Knudsen number $Kn(L) = \lambda/L$ is greater than 1, where $\lambda$ is the MFP of the gas and $L$ is a characteristic length scale of the structure for a particular photophoretic mechanism. For any given structure, $L$ depends on the photophoretic mechanism in question. For instance, thin structures that could levitate via $\Delta\alpha$ and $\Delta T$ photophoresis have $L$ equal to their smallest dimension normal to the force. The thin disks examined by Keith and Azadi et al., which levitate via $\Delta\alpha$ photophoresis, had $L$ equal to their diameters, thus limiting levitation to disks less than roughly 100 µm in diameter at the stratopause. Structures that can levitate via thermal transpiration, like nanocardboard [6,10], have interspaced holes or channels with smallest horizontal dimension equal to $L$ [11–13]. If the holes and channels satisfy $L \lesssim \lambda$, the only size limit on the entire levitating structure is set by the mechanical strength of the structure itself. We describe the simplest structure that could levitate using thermal transpiration in the methods section.

**Results and Discussion**

**Modeling.** Figure 1d shows a scanning electron microscope (SEM) image of a structure we wish to model. This alumina sandwich structure is divided into regions with and without vertical walls that connect the top and bottom face sheets. We fabricated the region with walls by directionally etching cylindrical channels into a silicon wafer, atomic layer depositing (ALD) 100 nm of alumina, then etching away the remaining silicon [7]. The region without walls was fabricated by etching holes in the ALD alumina instead of channels in the silicon. If lofted in the stratosphere, an absorptive coating on the bottom surface would heat the bottom layer relative to the top layer, creating a downward flow of air that thrusts the structure upward. We aim to study structures of order 10 cm wide, a feasible fabrication range as discussed in the practical design section.

Our one-dimensional analytical models represent a horizontally uniform bilayer structure with two infinitely thin, isothermal, perforated layers separated by a height $H$. The perforations have an area filling fraction, $f$, and the structural walls separating the layers have an area filling fraction $w$. The flow through the holes assumes they have dimensions much smaller than $\lambda$ to satisfy the FM regime criteria. Figure 1e shows a vertical cross-section of the generalized model structure labeled with heat resistivity terms.

At a distance from the structure equal to the thermal boundary layer thickness $d$, we assume the air has ambient temperature $T$. The distance $d$ is many orders of magnitude larger than $\lambda$ in the quiescent stratosphere. We assume uniform pressure (see Supplementary Information).

To calculate the thrust on the structure, we first solve a system of conservation of energy equations for $T_b$ and $T_t$, the structure's bottom and top layer temperatures, respectively (see methods section). We then solve for $T_{b,air}$ and $T_{t,air}$, the air temperatures one MFP below and above the structure (herein called "surface air layers"), respectively, by assuming linear temperature gradients $(T_b - T)/d$ and $(T_t - T)/d$ between the horizontal layers and the ambient air. We developed two analytical models valid in two different limits, $H \ll \lambda$ and $H \gg \lambda$. We could not develop an analytical model for $H \sim \lambda$ because the temperature gradient between the top and bottom layers is poorly defined in this range. The thrust equations for both models are functions of $T_b$, $T_t$, $T_{b,air}$, and $T_{t,air}$ and are given in the methods section. For the $H \gg \lambda$ model, we also calculate the air temperatures one MFP below and above the top and bottom layers, respectively, by assuming a linear vertical temperature gradient $(T_b - T_t)/H$ between the two layers.

The energy conservation equations contain the heat transfer terms shown in Fig. 1e. We compute the net radiative heat transfer to the top (bottom) layer by multiplying its solar band emissivity $\epsilon_{vis,t}$ ($\epsilon_{vis,b}$) and longwave thermal-IR band emissivity $\epsilon_{IR,t}$ ($\epsilon_{IR,b}$) by the incident radiative solar flux and terrestrial thermal upwelling, respectively. Initial calculations showed a roughly 1 % decrease in the thrust on the structure when accounting for a solar reflectivity of 0.05, typical of alumina [14]. For simplicity, we assume the structure is not reflective.

We also model the following five conductive heat transfer pathways, shown as resistors in Fig. 1e: between top and bottom layers via (1) the structural walls and (2) air in the adjacent hollow spaces, (3) between surface air layers via the air in the vertical regions where there are holes, and (4,5) from both sides of the structure to the environment via the thermal boundary layer. For each of these pathways, we account for the change in heat transfer coefficient as a function of pressure by summing the free-molecular and continuum heat transfer coefficients in parallel (equation (4) in the methods section), shown explicitly by two resistors in series for each case of boundary layer conduction in Fig. 1e. We assume the surface air layers have accommodation coefficients of 1.

We assume the same static boundary layer on both sides of the structure and do not explicitly calculate convective heat loss. The real boundary layer between a lofted structure and its surroundings may be poorly defined and will vary with time, especially in turbulence. This is the largest limitation of our models. Computational fluid dynamics (CFD) could characterize the boundary layer as a function of time, in more than one dimension, and on both sides of the structure.

To calculate $d$, we first assumed the structure was a disk with area density of order 1 g/m$^2$ falling at terminal velocity in the stratosphere. For a disk with diameter $D$ on the order of cm, $d$ is on the order of m (see Supplementary Information). This is unphysical since natural convection would limit the actual thermal boundary layer to a plume roughly the size of the structure's diameter. In fact, $d < D$ only when $D \gtrsim 10$ m. Figure 2a shows plots of $T_{b,air} - T_{t,air}$ and the daily time-averaged lofting force as $d$ varies parametrically. Increasing $d$ from 10 cm to 1 m reduces the time-averaged lofting force by only 0.5 %. The limit $d \to \infty$ maximizes the lofting force. The force varies by roughly 15 % in the range $d > 1$ mm. As $d$ decreases below this range, the lofting force drops significantly due to decreasing $T_{b,air} - T_{t,air}$. Given the insensitivity of the lofting force to macroscale boundary layer size, we set $d = D = 10$ cm for all results herein. This assumption gives a 15 % uncertainty in the lofting force and applies to the real macroscale structures we wish to model.

Our 1D models assume that the walls are uniform pillars spaced so closely that the effect of horizontal heat flow causing non-uniform temperatures in the top and bottom sheets can be neglected. For our engineering design, we developed a separate 2-dimensional heat flow model that considers thick walls separating large, horizontal, wall-free regions (see practical device design and methods sections). This 2D model allows us to compute the local reduction in temperature gradient between layers, and thus the lofting force, near walls.

Our 1D models apply to (1) structures with hole and wall patterns (i.e. unit cells) that are small compared to their macroscopic size, (2) nonturbulent environments with minimal airflow around a structure, and (3) stationary structures.

**Optimal parameters for maximum lofting forces.** Over the range of all parameters, the two 1D models converge when $H \sim \lambda$, or $Kn(H) \sim 1$. This value of $H$ maximizes the lofting force. Herein for any given structure, we only report the results of the $H \ll \lambda$ model for pressures below the intersection point and the results of the $H \gg \lambda$ model at higher pressures, effectively taking the minimum of both models at a given Knudsen number (Fig. 2b). The exact Knudsen number for maximum flow through a non-isothermal structure depends on the structure's geometry and properties of the gas. However, flows are typically maximized in the transition regime ($0.1 < Kn < 10$), usually around $Kn = 1$ [15,16]. For instance, Passian et al. [17] experimentally found the lofting force on a similar structure with $H = 2$ µm in air was maximized at $Kn(H) = 0.9$. Our results support these findings.

Figure 2c shows how the altitude of maximum lofting force varies with $H$. The maximum lofting force for a structure with $H = \lambda$ is roughly constant throughout the stratosphere, increasing 9 % from 20 km to 50 km. A structure can stably levitate if the time-averaged lofting force equals the gravitational force (the structure's weight) and if the time-averaged lofting force decreases with altitude. For instance, the structure described in Fig. 2b will levitate between 34 and 54 km, floating stably at 54 km, if its weight is 50 g/m$^2$. Structures with smaller $H$ are generally less stable in the stratosphere. For instance, because a 52 g/m² structure with $H = 2$ µm has such a broad range of float altitudes, it would need active navigational control to maintain an altitude of 25 km.

Figure 2d compares the mechanisms of heat transfer between horizontal layers. Air and structure conduction are the dominant mechanisms below 60 km. To find the value of the wall filling fraction $w$ at which the two are equal, we solve $k_{air} = w\, k_{mat}$, where $k_{air}$ is the conductivity of air between the top and bottom layers (given by equation (4) divided by $H$) and $k_{mat}$ is the conductivity of the structure material (in this case, 1.8 W m⁻¹ K⁻¹ for ALD alumina [18]). Air conduction dominates for $w < 10^{-3}$ and the lofting force is insensitive to $w < 10^{-4}$ (Supplementary Fig. 7). For comparison, nanocardboard has been fabricated with $w$ as low as 0.0018. An adequately stiff supporting structure could hold 100 nm thick alumina membranes a distance $H$ apart and without walls ($w = 0$) across horizontal distances on the cm scale [7]. Walls may only be needed to prevent the membranes from bending relative to each other and to provide adequate shear stiffness. Radiative heat transfer between sides is only dominant at upper mesospheric altitudes and heat transfer due to thermal transpiration is negligible at all altitudes.

The temperature difference $T_b - T_t$, and thus the thrust, is proportional to the difference in the net radiative fluxes on the top and bottom layers. Strict proportionality breaks down when thermal-infrared radiation (middle and long IR centered at 10 µm) becomes an important heat conduction term in the mesosphere.

Across a range of parameters, we found a hole filling fraction in the range $0.3 < f < 0.5$ maximizes the lofting force. Figure 2e shows the value $f = 0.47$ optimizes stratospheric lofting. This is close to the theoretical value of $f = 50\,\%$ calculated by Scandurra [16]. Both models approach this value as $H \to \lambda$. As $f \to 0$ and $f \to 1$, the lofting force approaches zero as expected.

Optimizing the optical properties of the structure to maximize lofting is straightforward in the solar spectral band: the thrust is maximized with a transmissive and nonabsorptive top layer ($\epsilon_{vis,t} = 0$) and absorptive bottom layer ($\epsilon_{vis,b} = 1$). In the thermal band, the optimal parameters depend on the time of day. During the daytime they are $\epsilon_{IR,b} = 0$ and $\epsilon_{IR,t} = 1$, corresponding to a perfect radiative cooling top layer and a perfect radiative warming bottom layer. At night, they are $\epsilon_{IR,t} = \epsilon_{IR,b} = 1$; a larger $\epsilon_{IR,b}$ increases the lofting force at night because terrestrial thermal upwelling is the only incident radiation. The optimal daytime emissivities maximize the time-averaged lofting force. A 10 cm diameter disk structure at an altitude of 25 km with $\epsilon_{vis,t} = \epsilon_{IR,b} = 0$, $\epsilon_{IR,t} = \epsilon_{vis,b} = 1$, $H = \lambda$ (equal to 2 µm at 25 km), $f = 0.47$, and $w = 0$, herein called our "benchmark parameters," has a time-averaged lofting force of 54 g/m², compared to 45 g/m² for a structure with emissivities optimized for nighttime levitation. To account for diurnal fluctuations, we weigh the solar flux by a positive sine wave with a period of 24 h. We assume lofting at the equatorial equinox; averaging over the year reduces the time-averaged lofting force by 3 %.

The lofting force generally increases in the stratosphere as the accommodation coefficient of either top ($\alpha_t$) or bottom ($\alpha_b$) surface decreases (Fig. 2f). When air conduction is the dominant heat transfer mechanism between layers, the heat flux between surfaces is proportional to equation (4). The flux is proportional to $\alpha_t$ and $\alpha_b$ in the limit $H \ll \lambda$ and is unaffected by $\alpha_t$ and $\alpha_b$ in the limit $H \gg \lambda$. Various materials and surface treatment techniques have achieved $\alpha$ near 0.2 [19]. For all calculations in this paper, however, we assumed a benchmark value of $\alpha_t = \alpha_b = 0.6$, which is low but common for a range of materials in air [20,21]. Assuming otherwise benchmark parameters at 25 km, setting $\alpha_t = \alpha_b = 0.2$ gives a time-averaged lofting force of 78 g/m², whereas setting $\alpha_t = \alpha_b = 1$ gives a force of 43 g/m². With $H = \lambda$, radiative heat transfer between the top and bottom structure layers equals heat conduction through air between the layers at 30 km when $\alpha = 0.05$. This value is improbably low for most materials, so air conduction will always dominate heat transfer between layers in the stratosphere if $w < 10^{-3}$.

**Practical device design**. A useful device must communicate (at least outward) and carry a payload. The ability to alter altitude and to fly horizontally are additional useful features particularly when there is sufficient control to allow station-keeping.

We now describe a reference design 10 cm in diameter that could levitate a payload of roughly 285 mg at 25 km with navigational control and 2-way communication with the ground. This device could be fabricated using existing techniques. We also describe advanced designs and hardware for both communication and attitude adjustment. Additional considerations regarding vertical and horizontal motion, deployment, and stratospheric exposure are discussed in the Supplementary Information.

The components of the device are (1) a photophoretically active structure (PAS), (2) a superstructure (SS) that supports sections or "cells" of the PAS, (3) a rigid, low aspect ratio "superduperstructure" (SDS) that prevents the device from buckling or fracturing during transport, deployment, and flight, (4) hardware for attitude adjustment and navigation, (5) hardware for communication, and (6) the sensing payload.

Consider the following reference design as a basis for analysis. The PAS (Fig. 3a) is a sandwich of two perforated 100 nm thick alumina sheets spaced 2 µm apart. The perforations (holes) have radii of 100 nm and account for 47 % of the area. Cylindrical support posts with 1 µm radii and 100 nm-thick walls work with the SS to maintain structural integrity of the PAS sandwich. The posts are patterned hexagonally across the PAS and spaced 60 µm apart; close enough to prevent problematic relative bending of the face sheets and sparse enough such that $w = 10^{-4}$. The 1 µm post radius is large enough for the posts to be fabricated using ALD but small enough to prevent the local reduction in $\Delta T$ from significantly reducing the total lofting force on the PAS. The sheets of the PAS are anchored to the SS (Figs. 3b and 3c), which is a honeycomb structure with 1 mm wide hexagons made from rectangular alumina tubes that are 1 mm tall, 100 µm wide, and made of 1.5 µm thick horizontal walls and 300 nm thick vertical walls.

The PAS and SS could be fabricated simultaneously using ALD on a silicon substrate (see Supplementary Information). ALD alumina is an excellent material for lofted structures due to its low thermal conductivity (1.8 W/m-K) and low solar and thermal-band emissivities (both < 0.1 for ~100 nm films) [18,22].

The bottom surface of the PAS is coated with a 200 nm thick multilayer of chromium and aluminum, which is highly solar-absorptive ($\epsilon_{vis,b} = 0.9$) and IR-transmissive ($\epsilon_{IR,b} < 0.05$) with area density 0.5 g/m² [23]. Solar-transmissive, IR-absorptive coatings are being developed for a range of applications, Perrakis et al. [24] developed a 850 nm thick solar-transmissive (> 0.99), IR-absorptive ($\epsilon_{IR,t} = 0.9$) multilayer weighing 2.1 g/m². Theoretically, much lighter (order 0.1 g/m²) coatings with comparable cooling efficacies could be fabricated [25]. We assume a more conservative top surface coating with $\epsilon_{IR,t} = 0.9$ and $\epsilon_{vis,t} < 0.05$ weighing 1 g/m².

At an altitude of 25 km near the equator, this PAS specification generates a time-averaged lofting force of 51 g/m². However, this does not account for heat flow through the SS which will reduce the temperature difference across the PAS near the SS structural members. We solved Poisson's equation for $\Delta T(x,y) = T_b(x,y) - T_t(x,y)$ (Fig. 4a) over a hexagonal $xy$ domain with boundary condition $\Delta T = 0$ to simulate perfect vertical heat conduction within the SS of temperature $T_w$ (see methods section). The results from one calculation are shown in Fig. 4b. We integrated $\Delta T$ over the domain and divided by the maximum attainable $\Delta T$ (assuming no SS or posts) to show that our reference design's 1 mm wide PAS cells reduce the lofting force by 5 % compared to the maximum (Fig. 4c).

We used the same approach to find the local reduction in lofting force near posts due to structural conduction (Fig. 4d). We imposed periodic boundary conditions around a hexagonal unit cell that defines the post spacing. Figure 4e shows the results from one calculation. Figure 4f shows that our reference design's post spacing of 60 µm ($w = 10^{-4}$) reduces the lofting force by 1 % compared to the maximum.

Figures 4c and 4e also show the lofting force is less sensitive to the PAS thickness $\delta$ for larger spacing between posts or SS walls. A thicker ALD alumina sandwich would increase the bending stiffness of the PAS cells. For PAS cells > 1 mm per side with post spacing > 60 µm, this would not meaningfully increase structural conduction nor increase the PAS weight relative to the payload capacity. However, it would decrease the fracture stress of the cells [7] and reduce the solar-band transmissivity of the top surface.

Together, our PAS and SS designs have an area density of 6.3 g/m$^2$, with 1.5 g/m$^2$ from coatings and 0.5 g/m$^2$ from the PAS alumina. Accounting for the reductions in the lofting force of 10 % from the SS area, 5 % from the 1 mm sided PAS cells, and 1 % from the 60 µm post spacing, the net time-averaged lofting force above the PAS and SS mass is 36.5 g/m$^2$.

Finite element analysis (FEA) shows 1 mm is a reasonable PAS cell width to withstand transportation, deployment, and flight without fracturing. Road transport and stratospheric balloon launches can involve up to $2g$ [26,27], where $g$ is gravitational acceleration. A 1 m/s wind normal to the device would create a Newtonian drag pressure of 0.3 Pa on the PAS. If deployed from a balloon, 1 m/s is a liberal estimate for the wind load a held structure may encounter as it becomes exposed to the open air. During flight, however, the PAS will need to lift the total weight of the device: roughly 40 g/m$^2$, 0.4 Pa, or $80g$ on the alumina sandwich.

We performed large deformation FEA on a 1 mm square face sheet of 100 nm alumina fixed on all sides (Fig. 5a) and a 1 mm-wide cantilever beam of our reference SS pattern fixed on one side (Fig. 5b). Assuming a fracture stress of 1 GPa, Young's modulus of 170 GPa, and Poisson's ratio of 0.21 for 100 nm-thick ALD alumina [28–30], a 1 mm sided PAS cell can withstand roughly 500 times the expected 0.4 Pa before fracturing when rigidly fixed on all sides. This PAS cell could also withstand $10^3$ times the critical buckling moment of the SS before fracturing (see Supplementary Information). Our reference PAS design may therefore seem overbuilt but it is more robust for handling during fabrication while only marginally decreasing the net lofting force of a practical device.

When fixed on one side with periodic boundary conditions perpendicular to its length, the SS cantilever beam can withstand at least 40 times its critical buckling moment before fracturing. In a practical device, the SS would be supported by an adequately stiff SDS such that vertical shear forces are not the primary failure mechanism of the SS. We therefore ignore shear stresses and post-buckling behavior and focus on buckling as the primary failure mode. Alternate designs with improved shear stiffness are discussed in the Supplementary Information. We found our SS design will buckle under its own weight if its total size is larger than a 12 cm-sided square (see Supplementary Information). Our reference SS should be stiff enough to manipulate during fabrication and deployment with only minimal added support from a macroscopic support frame or an SDS.

Kim et al. [7] explores the contributions of the face sheets and walls to the mechanical behavior of generalized nanocardboard plates; our PAS is one example. In that paper and in the Supplementary Information, we show the maximum stress expected in our reference PAS is insensitive to the presence of walls. Walls provide little bending stiffness when a force is applied normal to the face of the PAS and we do not expect significant shear stresses in the PAS due to the much higher bending stiffness of the SS. Posts need to be spaced < 500 µm apart to prevent the PAS from fracturing. This spacing violates the "no-straight-line rule," but still provides sufficient stiffness for expected loads [7,10].

In our reference design, the SDS is a space-filling truss that stiffens the SS. Space-filling trusses generally provide more mass-specific rigidity than a flat structure. With less mass than a thicker SS, the SDS prevents SS failure due to encountered loads, residual stresses, and fabrication defects. We designed an unoptimized truss with three 5 cm-long legs (Figs. 3d and 5c) in SkyCiv software. This 10 mg truss, made of carbon fiber reinforced plastic I-beams, supports roughly 10 times the maximum expected load on the entire device (see Supplementary Information). An unoptimized 230 mg truss with three 10 cm-long legs (Fig. 5c) could support a 20 cm diameter disk with the same strength. Larger devices with optimized trusses might achieve payloads of a few grams. SDSs could be 3D printed with a variety of materials, including reinforced plastics. While traditional 3D printers have adequate resolution (of order $10^3$ µm$^3$) to print SDSs, nanoscale direct laser writing (DLW) could create more efficient space-filling trusses made of reinforced polymers, ceramics, or metals with features on the scale of 100 nm and densities of order 100 kg/m$^3$ [31–33]. Regions of silicon retained on the bottom of an SS could be chemically bonded to an SDS.

In summary, our reference design is a 10 cm disk weighing 30 mg with a payload capacity of about 285 mg at 25 km. This design is not optimal. It is a compromise between the goals of maximizing payload, specifying a design that could be fabricated with existing methods, and analytical simplicity.

Alternate designs include an SDS shaped as a ring along the outside of the smaller superstructure (Fig. 3e). If this structure is a "supercell," then multiple supercells might be combine to form a larger honeycomb. To further support any SDS design, the PAS and SS could be fabricated with a dome-shaped profile (Fig. 3h). FEA showed a 200 µm wide, 20 µm tall half-spheroid structure has 75 % higher bending stiffness than a flat plate of the same thickness (Supplementary Fig. 8). The PAS and SS could be constructed atop dome-shaped falsework. A shallow dome profile minimizes both the horizontal component of the photophoretic force and any added weight relative to a flat disk.

**Attitude adjustment.** Concentrating the payload in the bottom center of the structure will passively align the structure upward via gravity, but we must actively control flight attitude to steer the device. The most plausible method is to attach a weight (or the bulk of the payload) to a rigid shaft hanging from the bottom center of the structure, like a pendulum. A micro electromechanical system (MEMS) actuator could change the hanging angle or horizontal position of the shaft to tilt the structure as needed (Fig. 3f). Milligram-scale electromagnetic and piezoelectric linear MEMS actuators can move hundreds of times their own weight [34–36] and could be coupled with rotational MEMS stages to more effectively angle the shaft [37,38]. The shaft could be made from the 3D printable materials discussed earlier. If $L$ is the length of the shaft (assumed to be weightless for simplicity) and $\theta$ is the angle between the shaft and vertical, the torque per unit weight at the end of the shaft is $Lg \sin \theta$, which is limited to below 1 Nm/kg for a 10 cm shaft length. This system would restore a vertically aligned structure to horizontal on the order of a second (the pendulum's natural period).

A purely active orientation method is to block local thermal transpiration using sliding plates attached to the PAS and with a matching hole pattern (Fig. 3g). The plates could slide to "open" or "close" holes in the PAS using linear MEMS actuators. Alternatively, we could change local temperature gradients. A domed structural profile does this passively with stronger lofting forces on areas with more solar insolation (Fig. 3h). Analysis in the Supplementary Information suggests that active attitude adjustment would allow our practical device to station-keep over a single ground station in some stratospheric regions and remain lofted within a vertical diurnal range of 700 m.

**Communication.** Lofted devices without the ability to communicate are useless for atmospheric sensing unless they can be recovered. While recovery seems implausible, communication is not. Assume the minimum setup required for communication includes a transceiving radio antenna, a perovskite or organic solar cell, and any necessary processing and regulating integrated circuits (ICs). Each component is much lighter than the 285 mg payload capacity of a 10 cm device. For instance, commercial amplifiers, microprocessors, and power regulation ICs on the order of 0.1 mg are readily available, and a 1 $cm^2$ solar cell of thickness 1 µm weighs roughly 0.2 mg. The antenna can be etched directly into the SS [41–43].

A device with these components could communicate continuously during the day. A 3 $cm^2$ solar cell with 20 % efficiency would source roughly 100 mW at peak daytime. An 80 % efficient, 10 cm (3 GHz), 0 dB gain antenna at 25 km can send on the order of 10 Mb/s to a 1 $m^2$ aperture on the ground with noise temperature 25 K and 10 dB signal-to-noise ratio (SNR). A 10 W ground transmitter would send data to the device with the same rate, gain, and SNR. Locating a lofted device requires transmitting into a large solid angle; once its position is known, the data rate can be increased by focusing the transmitter.

Communication at night requires onboard energy storage. Allocating roughly half our practical device's payload, 150 mg, to a supercapacitor with energy density 15 Wh/kg [44–46], the device could communicate 100 kb/s with SNR = 10 dB to the ground at night.

**Applications.** Stratospheric observation seems the most plausible near-term application. Nanoscale sensors have been developed for temperature, pressure, humidity, and radiometry [42,47,48]. A GPS receiver allows measurement of local stratospheric winds, which could improve weather prediction. Existing designs suggest it would be possible to develop a simple spectrally resolved solar-band sensor that scans zenith angles, perhaps to retrieve aerosol properties. Computation is not likely to be a limit; milligram integrated circuits (ICs) capable of sourcing their own power could easily integrate into practical structures [42]. Pairs of devices in formation flight could offer possibility of absorption measurements. Existing stratospheric *in situ* observations use aircraft such as the ER-2, which allow simultaneous measurement of chemical and physical properties along a flight track. A constellation of 100's of low-cost, long duration flying structures might compliment the limited aircraft observations despite having much less capable sensors.

Improved stratospheric sensing would be useful to monitor potential deployment of stratospheric aerosols for solar geoengineering (SG). Photophoretically lofted structures offer a few potential advantages as means of SG. They scatter light upwards eliminating the impact of SG on the visual appearance of the sky. They may allow more control of the latitudinal and spectral distribution of radiative forcing that is plausible with aerosols [4]. Macroscopic levitated structures are far too expensive for SG using current manufacturing techniques. Levitated structures for SG may be compared to space-based SG: both require manufacturing methods beyond current capabilities but may offer significant advantages over stratospheric aerosols.

Arrays of lofted structures could transmit or relay data. Daytime data rates can be high, a hundred structures that are slightly larger than our reference design could have daytime data rates comparable to the 10 Gb/s per satellite [49] for current low-earth orbit (LEO) satellite constellations.

Devices could be remotely illuminated with a ground-based near-IR laser and small (3-10 cm) aperture telescope enabling alternative alignment and navigation techniques. The lofting force on the structure could also be manipulated independent of attitude allowing a strong control of the structure's position. Remote illumination could power onboard components, allowing sensors to collect and transmit data at night with high rates [50].

Photophoretic devices could be deployed on other planets. On the surface of Mars, for instance, the time-averaged lofting force on a PAS with $H = \lambda = 10$ µm, $f = 0.37$, $\epsilon_{vis,t} = 0$, $\epsilon_{IR,t} = \epsilon_{vis,b} = \epsilon_{IR,b} = 1$, and otherwise reference parameters is 37 g/m², roughly equal to the lofting force in Earth's stratosphere, assuming ambient temperature 210 K and pressure 640 Pa. Dust storms might destroy practical structures at Mars's surface. Because storms are confined to the troposphere, however, lofted structures at the tropopause may have greater longevity. At the 40 km tropopause, a PAS with $f = 0.46$ and otherwise the same parameters would have a time-averaged lofting force of 18 g/m², assuming ambient temperature 156 K and pressure 10 Pa [51].

In summary, we present analytical models that calculate the photophoretic lofting force due to thermal transpiration on macroscopic bilayer sandwich structures. We used the models to find the optimal design, thermal, and optical properties of such a structure to maximize the lofting force. We analyze the model results to suggest photophoretic flight of a practical 10 cm diameter device in the stratosphere with capabilities for bi-directional communication with the ground and limited navigational control.

**Methods**

**Descriptions of symbols used in the main text**

| | |
|---|---|
| $F_p$ | photophoretic lofting force per unit area; positive in the upward direction |
| $T_b, T_t$ | temperatures of the bottom (top) layer of the structure |
| $T_{b,air}, T_{t,air}$ | temperatures of the bottom (top) surface air layer of the structure |
| $T_{bm}, T_{tm}$ | temperatures of the air layers 1 MFP above (below) the bottom (top) structural layer |
| $RF_b, RF_t$ | net radiative heat flux on the bottom (top) of the structure |
| $\rho_b, \rho_t$ | bottom (top) surface air layer density: for bottom calculated as $\rho_b = P/RT_{b,air}$ |
| $v_b, v_t$ | bottom (top) mean molecular gas speed; for bottom calculated as $v_b = \sqrt{2k_b T_{b,air}/(\pi m)}$ |
| $\alpha_b, \alpha_t$ | bottom (top) surface thermal accommodation coefficient, assumed to be 0.6 for all calculations except where noted |
| $f$ | hole filling fraction (fraction of the total horizontal area of the structure that is holes) |
| $w$ | wall filling fraction (fraction of the total internal volume of the structure that is wall material) |
| $H$ | structure height (distance between top and bottom surfaces) |
| $T$ | ambient temperature |
| $P$ | ambient pressure |
| $v$ | ambient mean gas speed |
| $d$ | thermal boundary layer thickness |
| $k_{air}$ | bulk thermal conductivity of air, 0.0284 W/(m K) |
| $k_{mat}$ | thermal conductivity of the scaffold material, 1.8 W/(m K) for ALD alumina [18] |
| $h_b, h_t,$ $h_m, h_{m,air}$ | heat transfer coefficients in air (W m$^{-2}$ K$^{-1}$) from the bottom layer to ambient, the top layer to ambient, between top and bottom layers, and between surface air layers, respectively |
| $\lambda$ | mean free path |
| $S(t)$ | solar constant, 1366 W/m$^2$, weighted by a positive half-sine wave with 24 h period |
| $A$ | average terrestrial albedo, 0.3 |
| $\epsilon_{vis,b}, \epsilon_{vis,t}$ | solar band emissivity of bottom (top) layer |
| $\epsilon_{IR,b}, \epsilon_{IR,t}$ | longwave (IR) band emissivity of the bottom (top) layer of the structure |
| $T_e$ | terrestrial thermal upwelling blackbody temperature, 255 K |
| $c_P$ | heat capacity of air, 1000 J/(kg K) |
| $\sigma$ | Stefan-Boltzmann constant, $5.67 \times 10^{-8}$ W/(m$^2$ K$^4$) |
| $k_B$ | Boltzmann constant, $1.38 \times 10^{-23}$ J/K |
| $R$ | ideal gas constant for dry air, 287.1 J/(kg K) |
| $m$ | mean molecular mass of air, $4.81 \times 10^{-26}$ kg |
| $\Delta T(x, y)$ | temperature difference between PAS layers, $T_b - T_t$, as a function of horizontal position $(x, y)$ |
| $T_w$ | temperature of superstructure walls |
| $\delta$ | thickness of PAS layers |

**Force due to thermal transpiration**

Consider an infinitely thin, nonconductive, horizontal plate surrounded by a gas of uniform pressure $P$, which has different temperatures above and below the plate. Suppose that the plate has periodically spaced holes with area filling fraction $f$ and radii much smaller than $\lambda$ ($Kn \ll 1$) so that any flow through the holes is rarefied. Suppose further that some external heat transfer process ensures the gas temperatures adjacent to the top and bottom of the plate are $T_{t,air}$ and $T_{b,air}$ respectively.

Thermal transpiration flow through the holes will create an upward force per unit area, i.e., a pressure, on the plate given by:
$$F_p = \bar{v} f (\rho_b v_b - \rho_t v_t) \tag{1}$$
where $\rho = P/RT_{air}$ is the air density, $v = \sqrt{2k_B T_{air}/(\pi m)}$ is the average molecular air speed, $R$ is the ideal gas constant, $k_B$ is the Boltzmann constant, $m$ is the average molecular mass of air, and the subscripts $t$ and $b$ denote the top and bottom of the plate [12]. The two terms in the parentheses of equation (1) are the mass air flows from the bottom to top regions and vice-versa, respectively, and $\bar{v} = (v_b + v_t)/2$.

In the free atmosphere the flow exerts a pressure on the plate. If instead the plate divides two regions of fixed volume, one has a "Knudsen pump," a device that produces an equilibrium pressure difference between the gas above and below the plate [2,52,53]. The pressure difference between adjacent chambers in a Knudsen pump is proportional to equation (1), with the proportionality determined by the shape of the barrier separating the chambers. Sharipov [11,53] found that rectangular channels connecting two chambers give pressure differences up to 80 % of that in equation (1).

**Thermal transpiration model equations**

To calculate the temperatures $T_b$ and $T_t$ of the model structure shown in Fig. 1e, we solve the following conservation of energy equations:

$$RF_b + RF_t = (T_b - T)h_b + (T_t - T)h_t \tag{2}$$

$$((RF_b - (T_b - T)h_b) - (RF_t - (T_t - T)h_t))(1 - f) = (T_b - T_t)\left(k_{mat}\frac{w}{H} + h_m(1 - f - w)\right)$$
$$+ (T_{b,air} - T_{t,air})\left(h_{m,air} + c_P(T_b - T_t)(\rho_b v_b - \rho_t v_t)\right) f \tag{3}$$

with

$$RF_b = S(t)\,\epsilon_{vis,b}\left(A + (1 - \epsilon_{vis,t})\right) + \epsilon_{IR,b}\,\sigma\left(T_e^{\,4} + \epsilon_{IR,t}T_t^4 - 2T_b^4\right)$$

$$RF_t = S(t)\,\epsilon_{vis,t}\left(1 + A(1 - \epsilon_{vis,b})\right) + \epsilon_{IR,t}\,\sigma\left((1 - \epsilon_{IR,b})T_e^{\,4} + \epsilon_{IR,b}T_b^4 - 2T_t^4\right)$$

$$h_b = \left(\left(\frac{k_{air}}{d}\right)^{-1} + \left(\frac{P\,v\,\alpha_b}{T}\right)^{-1}\right)^{-1}$$

$$h_t = \left(\left(\frac{k_{air}}{d}\right)^{-1} + \left(\frac{P\,v\,\alpha_t}{T}\right)^{-1}\right)^{-1}$$

$$h_m = \left(\left(\frac{k_{air}}{H}\right)^{-1} + \left(\frac{P\,v\,\alpha_b}{T}\right)^{-1} + \left(\frac{P\,v\,\alpha_t}{T}\right)^{-1}\right)^{-1}$$

$$h_{m,air} = \left(\left(\frac{k_{air}}{H + 2\lambda}\right)^{-1} + 2\left(\frac{P\,v}{T}\right)^{-1}\right)^{-1}$$

All heat conduction terms in the above equations have the general form $k\,\Delta T/l$, where $k$ is a conductivity, $\Delta T$ is a temperature difference, and $l$ is the length scale describing $\Delta T$. All conduction terms involving either structural layer has a heat transfer coefficient $k/l$ of the form

$$\left(\left(\frac{k_{air}}{l}\right)^{-1} + \left(\frac{P\,v\,\alpha}{T}\right)^{-1}\right)^{-1} \tag{4}$$

where $k_{air}$ is the bulk conductivity of air and $\alpha$ is the accommodation coefficient of the corresponding horizontal layer. The term on the left dominates when $l \gg \lambda$ and the other dominates when $l \ll \lambda$; the total heat transfer term is equivalent to summing two thermal resistors in series [54].

Equation (2) describes conservation of energy for the entire system, with the energy input being the sum of the net radiative forcing terms (left side) and the energy output being the sum of the thermal conduction terms away from the structure to the ambient atmosphere (right side).

Equation (3) describes conservation of energy among the top and bottom structural layers and top and bottom surface air layers. The left side describes the difference in losses to the environment between the structural layers. The right side's first term describes thermal conduction between the two layers vertically through, respectively, regions with walls and the adjacent hollow regions with solid top and bottom layers. Since $w$ also represents the area fraction of the structure containing walls, the term $(1 - f - w)$ describes the area fraction of the structure that has neither holes nor walls; in other words, where there is an air gap between the top and bottom layers. Equations containing this term are accurate for $w \ll f$ and $w + f < 1$, which are true for the structures described in this paper. The right side's second term has two parts that describe energy flow between surface air layers: (1) The left part describes conduction between the surface air layers. Accommodation coefficients are only defined for solids and liquids (not layers of gas) because solids and liquids are thermal reservoirs compared to incident gas molecules. Our models assume the surface air layers are horizontally uniform in temperature; thermal reservoirs by definition. Therefore, we assume the surface air layers have accommodation coefficients of 1, as shown in the term $h_{m,air}$. (2) The right part describes energy flow due to thermal transpiration, which is insignificant (Fig. 2d) and was therefore ignored when solving for the structure's temperatures. The net radiative heat fluxes $RF$ assume no reflection of incident light.

Next, we assume a linear temperature gradient between the structure and its surroundings to calculate $T_{b,air} = T_b - \lambda \frac{T_b - T}{d}$ and $T_{t,air} = T_t - \lambda \frac{T_t - T}{d}$. In the limit $H \ll \lambda$, the lofting force is calculated using equation (1). In the limit $H \gg \lambda$, the lofting force is calculated by first calculating the temperatures of the air layers one MFP above the bottom surface, $T_{bm}$, and one MFP below the top surface, $T_{tm}$, assuming a linear temperature gradient between the two sides of the structure: $T_{bm} = T_b - \lambda \frac{T_b - T_t}{H}$ and $T_{tm} = T_t + \lambda \frac{T_b - T_t}{H}$. The lofting force is then

$$F_p = f \left( \frac{v_b + v_{bm}}{2} (\rho_b v_b - \rho_{bm} v_{bm}) + \frac{v_t + v_{tm}}{2} (\rho_{tm} v_{tm} - \rho_t v_t) \right) / 2. \tag{5}$$

As discussed in the results section, the lofting force at all pressures and structure heights is assumed to be the minimum of the two $F_p$ values returned by equations (1) and (5). Equation (1) describes the low pressure, small $H$ regime and equation (5) describes the high pressure, large $H$ regime. Note that the $H \gg \lambda$ model does not explicitly conserve mass between sides. The ratio of flow through the top layer to flow through the bottom layer is $> 99$ % for a structure with optimal parameters in the stratosphere (Supplementary Fig. 6).

**The effect of structure conduction on horizontal temperature gradients in the PAS**

**Structure conduction through the SS.** Consider the temperature difference $\Delta T \equiv T_b - T_t$ between a PAS's two layers as a function of position in the horizontal $xy$ plane. Let the two layers have thickness $\delta$ and let an adjacent SS have uniform temperature $T_w$ and infinite thermal conductivity, so that $T_b = T_t = T_w$ at the junction of the superstructure and PAS (Fig. 3a). For simplicity, we ignore any effects from holes and the boundary layer. Poisson's equations on a bound domain in the $xy$ plane are

$$RF_t - \delta\, k_{mat}\, \nabla^2 T_t(x,y) + h_m\big(T_b(x,y) - T_t(x,y)\big) = 0$$
$$RF_b - \delta\, k_{mat}\, \nabla^2 T_b(x,y) + h_m\big(T_t(x,y) - T_b(x,y)\big) = 0.$$

This is a difficult system to solve because of the fourth order terms in $RF_b$ and $RF_t$. To linearize the system, we reduce it to one equation of $\Delta T$ by assuming $T_b(x,y) = T_w + \Delta T(x,y)/2$ and $T_t(x,y) = T_w - \Delta T(x,y)/2$. Raising these terms to the fourth power and expanding, ignoring second order and higher terms of $\Delta T$ because $\Delta T \ll T_w$, gives

$$T_b^4(x,y) \approx T_w^4 + 4T_w^3\, \Delta T(x,y)$$
$$T_t^4(x,y) \approx T_w^4 - 4T_w^3\, \Delta T(x,y).$$

We solved
$$RF_b - RF_t - \delta \nabla^2(\Delta T(x,y)) + h_m \Delta T(x,y) = 0$$
over a hexagonal domain shown in Fig. 4b with boundary condition $\Delta T(x,y) = 0$. We used benchmark daytime parameters at 25 km. We integrated the results over the domain area and divided the result by the maximum value of $\Delta T$ attained from a calculation where there was no structural conduction anywhere. The results are shown in Fig. 4c.

**Structure conduction through cylindrical posts.** Within the PAS, posts (walls) spaced far from holes maximize $\Delta T$ near the edges of the holes. We wished to find the reduction in the local $\Delta T$ near cylindrical posts. Using the same approach as above, we defined posts in the shape of an open cylinder with 1 µm radius and conductivity $k_{mat} = 1.8$ W/(m K) centered at the origin and at each of the corners of a hexagonal PAS unit cell. We did not consider any SS effects and we assumed periodic boundary conditions along the edge of the domain. We used benchmark daytime parameters at 25 km. We again integrated the results over the domain area and divided the result by the maximum value of $\Delta T$ to produce the results shown in Fig. 4f.

Posts with infinite structure conduction (i.e. $\Delta T = 0$ at their perimeters) have nonzero thermal transpiration flow through their openings if their radii are $< \lambda$. We do not consider this effect in the above analysis because the post radii are roughly equal to $\lambda$ at 25 km and the area filling fraction of post holes is insignificant ($< 1$ %) at a post spacing of 60 µm. We discuss this effect in the Supplementary Information.

**Acknowledgments**

We thank S. Catsoulis, J. Dykema, A. Feldhaus, M. Greenberg, J.F.M. Hardigree, P. Horowitz, Q. Jiao, F. Keutch, Y. Shyur, J. Smith, and L. Wai for fruitful discussions. This work was supported by the Star Friedman Challenge for Promising Scientific Research at Harvard University and the Harvard University Solar Geoengineering Research Program.

**Author Contributions**

D.W.K. and J.J.V. designed the research. B.C.S. and D.W.K developed the models. B.C.S. wrote and ran the models. J.J.V. and J.K. developed the PAS and SS design and fabrication process. J.K. performed FEA. B.C.S. and D.W.K. wrote the paper.

**Competing interests statement**

The authors declare no competing interests.

**References**

1. Smalley, D. E. *et al.* A photophoretic-trap volumetric display. *Nature* **553**, 486–490 (2018).
2. Gupta, N. K. & Gianchandani, Y. B. Thermal transpiration in zeolites: A mechanism for motionless gas pumps. *Appl. Phys. Lett.* **93**, 193511 (2008).
3. Redding, B., Hill, S. C., Alexson, D., Wang, C. & Pan, Y.-L. Photophoretic trapping of airborne particles using ultraviolet illumination. *Opt. Express* **23**, 3630 (2015).

4. Keith, D. W. Photophoretic levitation of engineered aerosols for geoengineering. *Proc. Natl. Acad. Sci. U.S.A.* **107**, 16428–16431 (2010).
5. Azadi, M. *et al.* Controlled levitation of nanostructured thin films for sun-powered near-space flight. *Sci. Adv.* **7**, eabe1127 (2021).
6. Cortes, J. *et al.*, Photophoretic levitation of macroscopic nanocardboard plates. *Adv. Mater.* **32**, 1906878 (2020).
7. Kim, J. *et al.* Ultralight and ultra-stiff nano-cardboard panels: Mechanical analysis, characterization, and design principles. Preprint at https://arxiv.org/abs/2211.02803 (2022).
8. Horvath, H. Photophoresis – a forgotten force ?? *KONA Powder Part. J.* **31**, 181–199 (2014).
9. Reynolds, O. On certain dimensional properties of matter in the gaseous state. *Philos. Trans. R. Soc.* **170**, 727–845 (1879).
10. Lin, C. *et al.* Nanocardboard as a nanoscale analog of hollow sandwich plates. *Nat. Commun.* **9**, 4442 (2018).
11. Sharipov, F. Non-isothermal gas flow through rectangular microchannels. *J. Micromech. Microeng.* **9**, 394–401 (1999).
12. Sharipov, F. & Seleznev, V. Data on internal rarefied gas flows. *J. Phys. Chem. Ref. Data* **27**, 657–706 (1998).
13. Han, Y.-L, Phillip Muntz E., Alexeenko, A. & Young, M. Experimental and computational studies of temperature gradient–driven molecular transport in gas flows through nano/microscale channels. *Nanoscale Microscale Thermophys. Eng.* **11**, 151–175 (2007).
14. Kumar, P., Wiedmann, M. K., Winter, C. H. & Avrutsky, I. Optical properties of $Al_2O_3$ thin films grown by atomic layer deposition. *Appl. Opt.* **48**, 5407 (2009).
15. Rohatschek, H. Semi-empirical model of photophoretic forces for the entire range of pressures. *J. Aerosol Sci.* **26**, 717–734 (1995).
16. Scandurra, M. Enhanced radiometric forces. Preprint at https://arxiv.org/abs/physics/0402011 (2004).
17. Passian, A., Warmack, R. J., Ferrell, T. L. & Thundat, T. Thermal Transpiration at the Microscale: A Crookes Cantilever. *Phys. Rev. Lett.* **90**, 124503 (2003).
18. Cappella, A. *et al.* High temperature thermal conductivity of amorphous $Al_2O_3$ thin films grown by low temperature ALD. *Adv. Eng. Mater.* **15**, 1046–1050 (2013).
19. Honig, C. D. F. & Ducker, W. A. Effect of Molecularly-Thin Films on Lubrication Forces and Accommodation Coefficients in Air. *J. Phys. Chem. C* **114**, 20114–20119 (2010).
20. Wachman, H. Y. The thermal accommodation coefficient: a critical survey. *J. Am. Rocket Soc.* **32**, 1 (1962).
21. Angelo, R. L. Experimental determination of selected accommodation coefficients. (California Institute of Technology, 1951).
22. Koushki, E., Mousavi, S. H., Jafari Mohammadi, S. A., Majles Ara, M. H. & de Oliveira, P. W. Optical properties of aluminum oxide thin films and colloidal nanostructures. *Thin Solid Films* **592**, 81–87 (2015).
23. Li, Z., Palacios, E., Butun, S., Kocer, H. & Aydin, K. Omnidirectional, broadband light absorption using large-area, ultrathin lossy metallic film coatings. *Sci. Rep.* **5**, 15137 (2015).
24. Perrakis, G. *et al.*, Submicron organic–inorganic hybrid radiative cooling coatings for stable, ultrathin, and lightweight solar cells. *ACS Photonics*, **9**, 4, 1327–1337 (2022).
25. Ilic, O. & Atwater, H. A. Nanophotonic heterostructures for efficient propulsion and radiative cooling of relativistic light sails. *Nano Lett.* **18**, 9, 5583–5589 (2018).

26. Sandia Laboratories. *Shock and vibration environments for large shipping containers on rail cars and trucks*. (1977).
27. Columbia Scientific Balloon Facility. *Structural requirements and recommendations for balloon gondola design*. (2013).
28. Ruoho, M. *et al.* Thin-Film Engineering of Mechanical Fragmentation Properties of Atomic-Layer-Deposited Metal Oxides. *Nanomaterials* **10**, 558 (2020).
29. Rontu, V. *et al.* Elastic and fracture properties of free-standing amorphous ALD $Al_2O_3$ thin films measured with bulge test. *Mater. Res. Express* **5**, 046411 (2018).
30. Berdova, M. *et al.* Fracture properties of atomic layer deposited aluminum oxide free-standing membranes. *J. Vac. Sci. Technol. A* **33**, 01A106 (2015).
31. Crook, C. *et al.* Plate-nanolattices at the theoretical limit of stiffness and strength. *Nat. Commun.* **11** (2020).
32. Jang, D., Meza, L. R., Greer, F. & Greer, J. R. Fabrication and deformation of three-dimensional hollow ceramic nanostructures. *Nat. Mater.* **12**, 893–898 (2013).
33. Bauer, J., Hengsbach, S., Tesari, I., Schwaiger, R. & Kraft, O. High-strength cellular ceramic composites with 3D microarchitecture. *Proc. Natl. Acad. Sci. U.S.A.* **111**, 2453–2458 (2014).
34. Algamili, A. S. *et al.*, A review of actuation and sensing mechanisms in MEMS-based sensor devices. *Nanoscale Res. Lett.* **16**, 16 (2021).
35. Acerce, M., Akdoğan, E. K. & Chhowalla, M. Metallic molybdenum disulfide nanosheet-based electrochemical actuators. *Nature* **549**, 370–373 (2017).
36. Ruiz-Díez, V. *et al.* Piezoelectric MEMS Linear Motor for Nanopositioning Applications. *Actuators* **10**, 36 (2021).
37. Baran, U. *et al.* Linear-stiffness rotary MEMS stage. *J. Microelectromech. Syst.* **21**, 514–516 (2012).
38. Gokce, S. K. *et al.* "2D scanning MEMS stage integrated with microlens arrays for high-resolution beam steering" in *2009 IEEE/LEOS International Conference on Optical MEMS and Nanophotonics*, (IEEE, 2009), pp. 43–44.
39. Lang, F., Wang, H., Zhang, S., Liu, J. & Yan, H. Review on Variable Emissivity Materials and Devices Based on Smart Chromism. *Int. J. Thermophys.* **39**, 6 (2017).
40. Athanasopoulos, N. & Siakavellas, N. J. Programmable thermal emissivity structures based on bioinspired self-shape materials. *Sci. Rep.* **5**, 17682 (2015).
41. Bharadwaj, P., Deutsch, B. & Novotny, L. Optical Antennas. *Adv. Opt. Photon.* **1**, 438 (2009).
42. Cortese, A. J. *et al.* Microscopic sensors using optical wireless integrated circuits. *Proc. Natl. Acad. Sci. U.S.A.* **117**, 9173–9179 (2020).
43. Zhou, W., Suh, J. Y. & Odom, T. W. Novel Fabrication Methods for Optical Antennas in *Optical Antennas* (eds. Agio, M. & Alu, A.) 277–293 (Cambridge University Press, 2012).
44. Nalwa, K. S. & Chaudhary, S. Design of light-trapping microscale-textured surfaces for efficient organic solar cells. *Opt. Express* **18**, 5168 (2010).
45. Alves, M., Pérez-Rodríguez, A., Dale, P. J., Domínguez, C. & Sadewasser, S. Thin-film micro-concentrator solar cells. *J. Phys. Energy* **2**, 012001 (2019).
46. Tyagi, A., Tripathi, K. M. & Gupta, R. K. Recent progress in micro-scale energy storage devices and future aspects. *J. Mater. Chem. A* **3**, 22507–22541 (2015).
47. Wu, X. *et al.* A $0.04mm^3$ 16nW wireless and batteryless sensor system with integrated Cortex-M0+ processor and optical communication for cellular temperature measurement. *2018 IEEE Symposium on VLSI Circuits* (IEEE, 2018).

48. He, P. *et al.* Fully printed high performance humidity sensors based on two-dimensional materials. *Nanoscale* **10**, 5599–5606 (2018).
49. Asian Development Bank. *Digital Connectivity and Low Earth Orbit Satellite: Constellations Opportunities for Asia and the Pacific*. (2021).
50. Mita, Y. *et al.* Microscale ultrahigh-frequency resonant wireless powering for capacitive and resistive MEMS actuators. *Sens. Actuator A Phys.* **275**, 75–87 (2018).
51. Petropoulos, B. & Macris, C. Physical parameters of the martian atmosphere. *Earth Moon Planets* **46**, 1–30 (1989).
52. Wang, X., Su, T., Zhang, W., Zhang, Z. & Zhang, S. Knudsen pumps: a review. *Microsyst. & Nanoeng.* **6**, 1–28 (2020).
53. An, S., Gupta, N. K. & Gianchandani, Y. B. A Si-Micromachined 162-Stage Two-Part Knudsen Pump for On-Chip Vacuum. *J. Microelectromech. Syst.* **23**, 406–416 (2014).
54. Trott, W. M., Castañeda, J. N., Torczynski, J. R., Gallis, M. A. & Rader, D. J. An experimental assembly for precise measurement of thermal accommodation coefficients. *Rev. Sci. Instrum.* **82**, 035120 (2011).

**Figures**

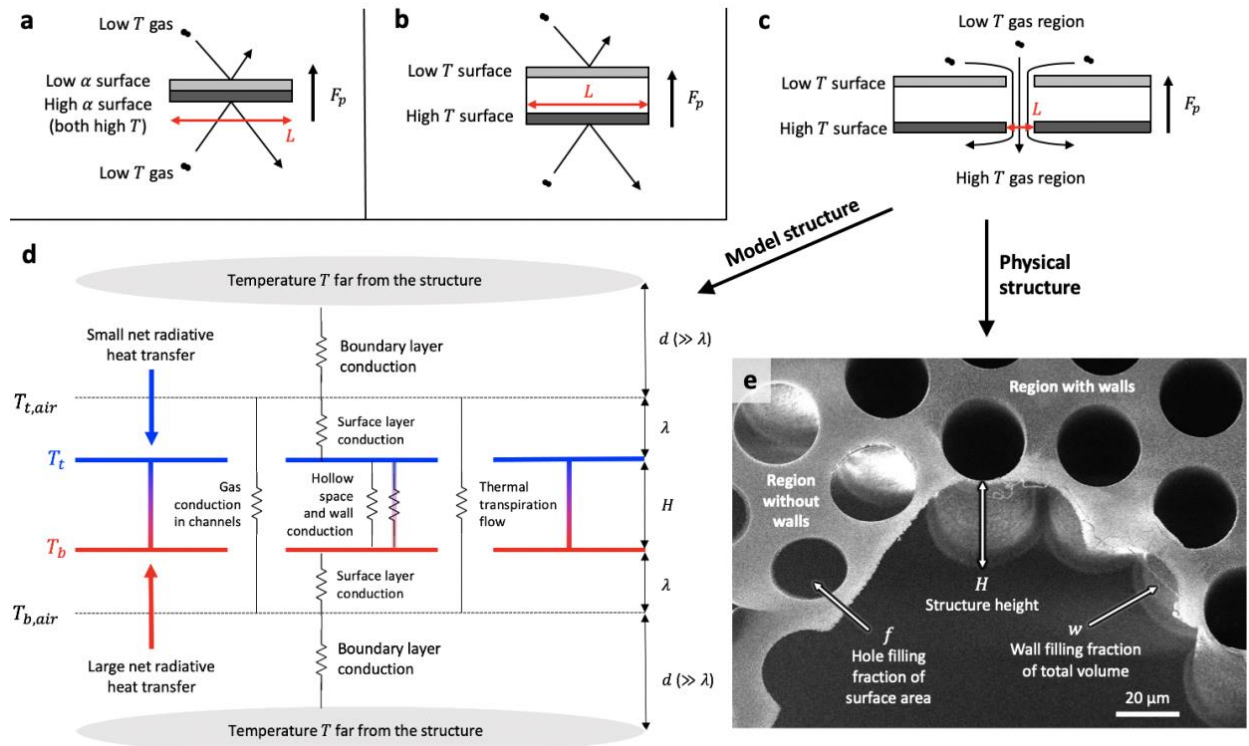

**Figure 1.** The three photophoretic mechanisms which produce an upward lofting force on an object when a characteristic length scale $L$ is less than the MFP (**a**-**c**) and physical and model structures that may produce a lofting force via thermal transpiration (**d** and **e**). **a** $\Delta\alpha$ photophoresis produced by a difference in $\alpha$ between the top and bottom surfaces of an isothermal structure that is warmer than the surrounding gas. **b** $\Delta T$ photophoresis produced by a difference in temperature between the top and bottom surfaces. **c** Thermal transpiration caused by the difference in air temperature above and below the structure. In **a** and **b**, $L$ is the minimum horizontal dimension of the structure, whereas in **c** $L$ is the horizontal dimension of the hole or channel. **d** SEM image of an example structure made of 100 nm thick alumina, showing the structure height $H = 100$ µm, wall filling fractions $w = 0.006$ in the region with walls and $w = 0$ elsewhere, and hole filling fraction $f = 0.4$. Holes were not etched in bottom layer of the structure for clarity. The region with walls varies from our model structure because the walls are concentric with the holes, which reduces the local temperature gradient $T_b - T_t$ and the resulting thermal transpiration flow at each hole. **e** Generalized heat resistivity diagram for the 1D models. Each resistor represents a heat transfer term in the models.

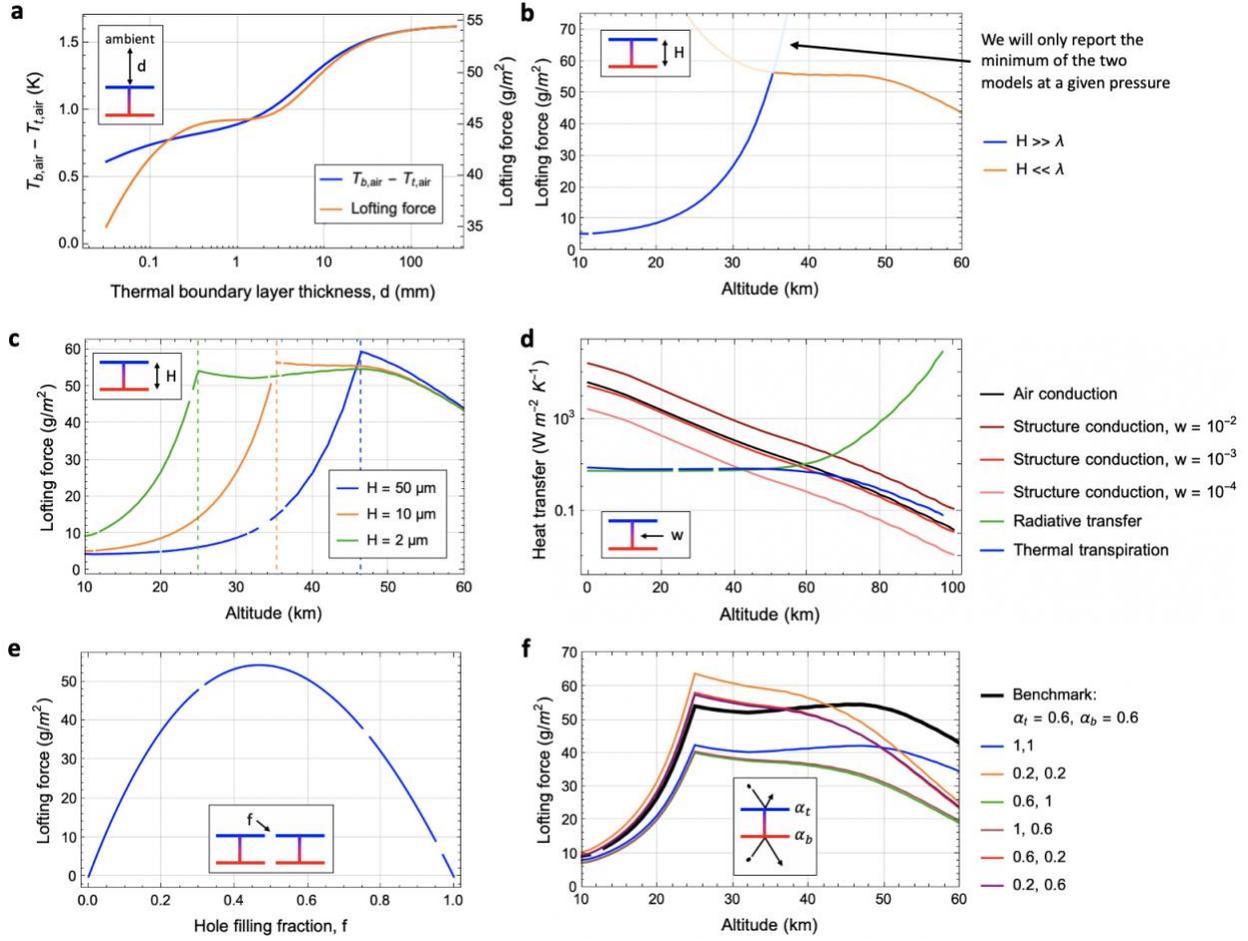

**Figure 2.** Analytical model results. Unless otherwise specified, all plots assume a structure with $\alpha_t = \alpha_b = 0.6$, $f = 0.47$, $w = 0$, $\epsilon_{vis,t} = \epsilon_{IR,b} = 0$, $\epsilon_{IR,t} = \epsilon_{vis,b} = 1$, and $d = 10$ cm. The insets depict the structural parameters of interest to each plot. **a** Plots of daytime $T_{b,air} - T_{t,air}$ and the time-averaged (over the diurnal cycle) lofting force as $d$ varies parametrically, with $H = \lambda = 2$ µm at 25 km. **b** Time-averaged lofting forces vs. altitude for both $H \ll \lambda$ and $H \gg \lambda$ models. We use the minimum of the two models at a given $Kn(H)$ as an estimate of the lofting force. **c** Time-averaged lofting forces vs. altitude for structures with $H = 2$, 10, and 50 µm. Respectively, the vertical dashed lines correspond to the altitudes of 25 km, 35 km, and 46 km where $Kn(H) = 1$. **d** Comparison of heat transfer terms vs. altitude for a structure with $H = \lambda$. For structure conduction ($w > 0$), we assume walls are made of ALD alumina with thermal conductivity 1.8 W m$^{-1}$ K$^{-1}$. **e** Time-averaged lofting force vs. $f$ with $H = \lambda = 2$ µm at 25 km. **f** Time-averaged lofting force vs. altitude for structures with several different values of $\alpha_t$ and $\alpha_b$ with $H = \lambda$.

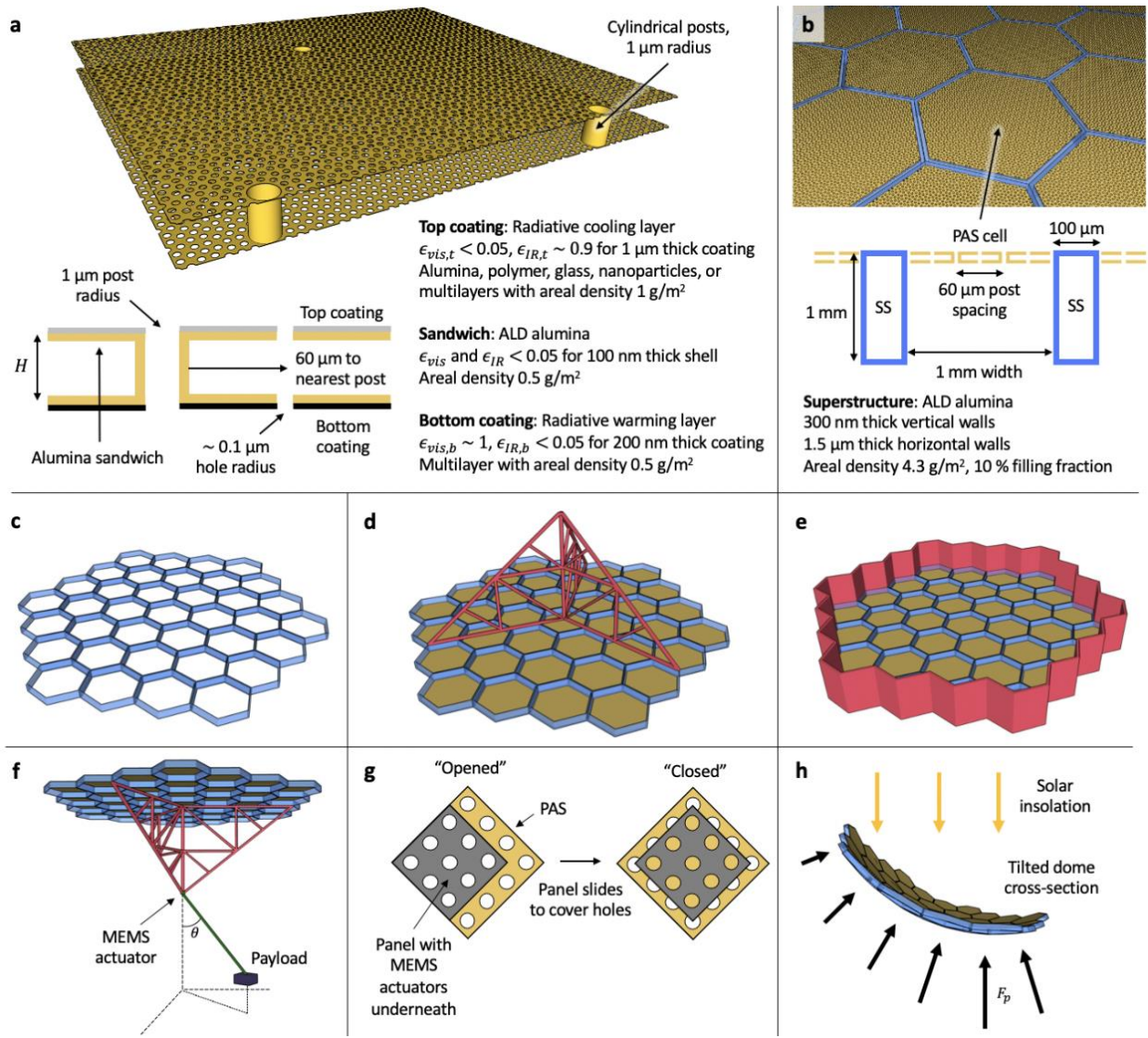

**Figure 3.** The structural components of a practical levitating structure (**a-e**) and three possible attitude adjustment methods (**f-h**). **a** An example photophoretically active structure (PAS, not to scale) and a vertical cross-section with coatings. **b** Hexagonal cells of PAS (gold) supported by a honeycomb superstructure (SS, blue) and a vertical cross-section. The PAS cells are flush with the top of the honeycomb lattice. **c** SS honeycomb lattice without PAS cells. **d** Bottom-up view of a SDS (red) in the form of a truss network. **e** Top-down view of an SDS in the form of a ring along the perimeter of the honeycomb network. **f** A payload attached to a rigid shaft and angled with a MEMS actuator to adjust the tilt angle. **g** Panels with the same hole pattern as the PAS could slide to block local thermal transpiration using linear MEMS actuators. **h** Time-averaged insolation creates a restoring torque on a dome-shaped structure tilted away from the horizontal.

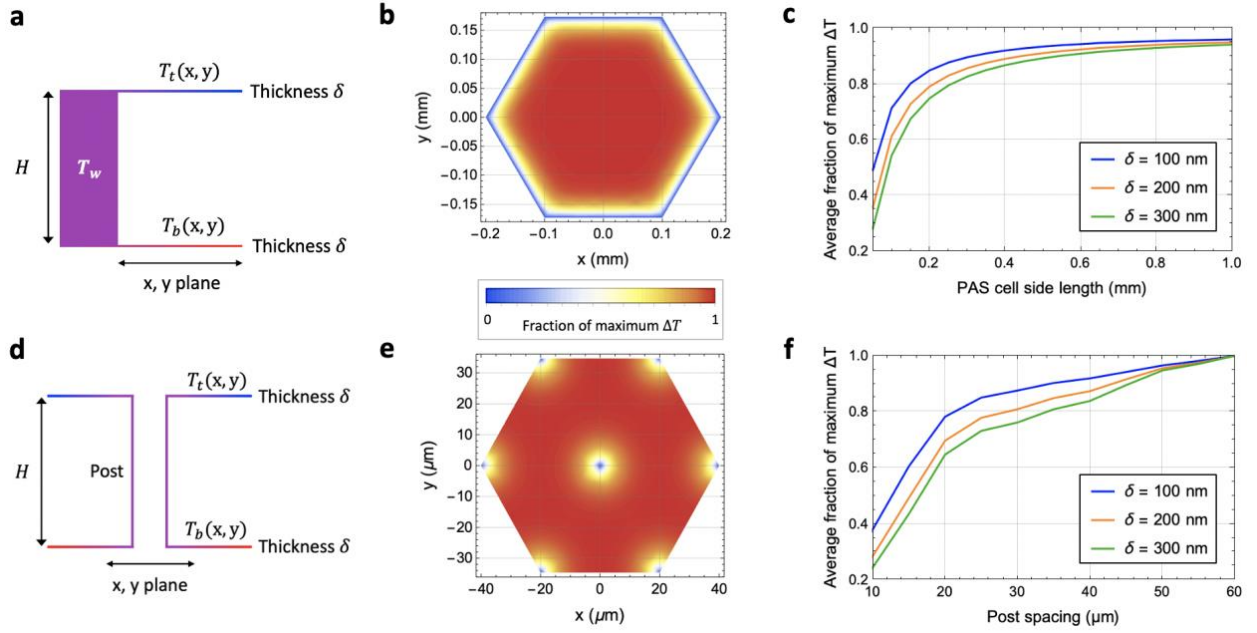

**Figure 4.** Setup and results from 2D heat models applied to the PAS. **a** Side view of a junction between a PAS cell with layers of temperatures $T_b$ and $T_t$ and infinitely conductive SS of temperature $T_w$. The horizontal layers have thickness $\delta$. **b** $\Delta T = T_b - T_t$ plotted over a hexagonal domain representing a model PAS cell, assuming benchmark parameters at 25 km. We show a hexagonal side length of 0.2 mm to illustrate the transition from $\Delta T = 0$ at the boundaries to its maximum value. **c** Plots of the ratio of average value of $\Delta T$ on the hexagonal domain to the maximum value of $\Delta T$ (infinitely far from the SS) as a function of PAS cell side length for different face sheet thicknesses $\delta$. **d** Side view of a cylindrical post separating the top and bottom PAS layers. The post walls and the horizontal layers have the same thickness $\delta$. **e** $\Delta T$ plotted over a domain representing a unit cell within the PAS that defines the hexagonal post spacing. Posts are located at the center and all corners of the hexagonal unit cell and the unit cell side length is the post spacing. We assumed benchmark parameters at 25 km and show a side length of 20 µm to illustrate the transition from $\Delta T \approx 0$ at the posts to its maximum value. **f** Plots of the ratio of average value of $\Delta T$ on the hexagonal domain to the maximum value of $\Delta T$ (infinitely far from posts) as a function of post spacing for different $\delta$ values.

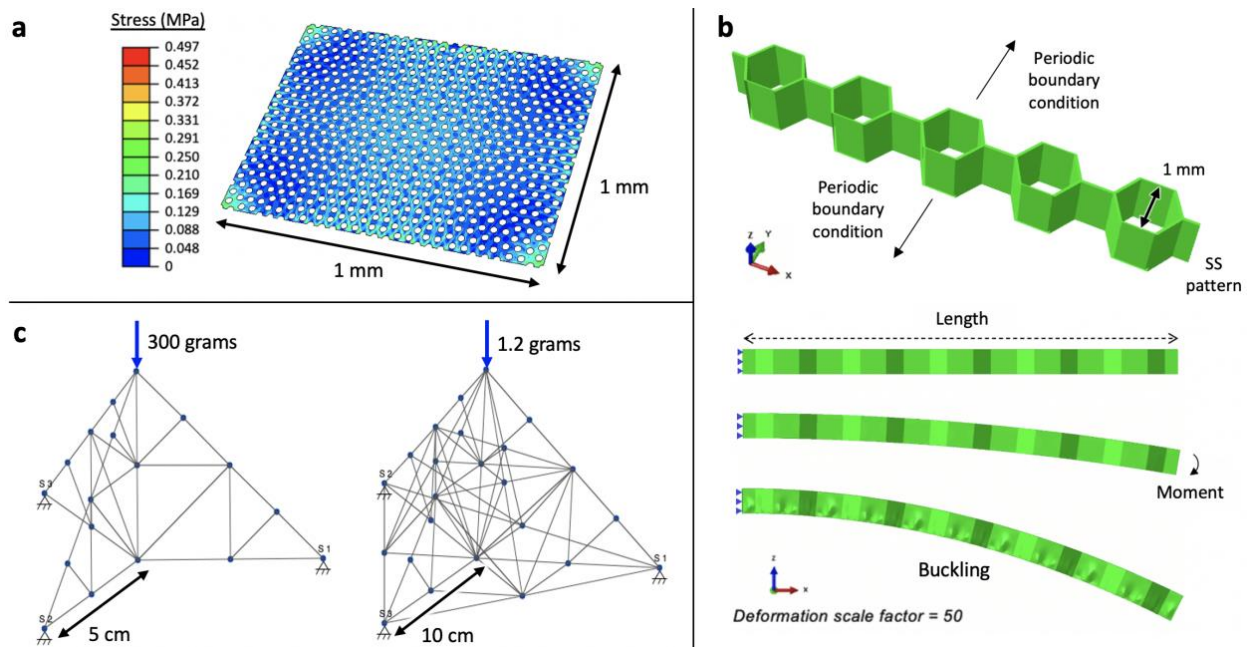

**Figure 5.** FEA and beam calculations used to find PAS, SS, and SDS designs that would withstand the expected forces of transport, deployment, and flight. **a** Stress map of a single 1 mm by 1 mm PAS face sheet without walls or posts when fixed on all sides and subject to 0.4 Pa. **b** Setup and results of FEA on a 0.52 mm-sided honeycomb SS pattern without PAS cells. The beam at top left is fixed at one end and a bending moment is applied on the other end. Periodic boundary conditions are applied in the $\pm y$ direction. Buckling occurs when the flexural rigidity of the cantilever drops suddenly as the applied moment increases (see Supplementary Information). **c** Unoptimized truss designs as SDSs for 10 cm (left) and 20 cm (right) diameter devices viewed from the side. The leg lengths and height are 5 and 10 cm, respectively. The trusses are pinned on 3D-hinges at the bases of their legs.

Supplementary Information for

# Analytical models for the design of photophoretically levitating macroscopic sensors in the stratosphere

Benjamin C. Schafer[1,*], Jong-hyoung Kim[1], Joost J. Vlassak[1], David W. Keith[1]

[1]John A. Paulson School of Engineering and Applied Sciences, Harvard University, Cambridge, MA 02138

*Email: schafer@g.harvard.edu

## Supplementary Text

### Descriptions of symbols used in the main text and Supplementary Information

| | |
|---|---|
| $F_p$ | photophoretic lofting force per unit area; positive in the upward direction |
| $T_b, T_t$ | temperatures of the bottom (top) layer of the structure |
| $T_{b,air}, T_{t,air}$ | temperatures of the bottom (top) surface air layer of the structure |
| $T_{bm}, T_{tm}$ | temperatures of the air layers 1 MFP above (below) the bottom (top) structural layer |
| $RF_b, RF_t$ | net radiative heat flux on the bottom (top) of the structure |
| $\rho_b, \rho_t$ | bottom (top) surface air layer density: for bottom calculated as $\rho_b = P/RT_{b,air}$ |
| $v_b, v_t$ | bottom (top) mean molecular gas speed; for bottom calculated as $v_b = \sqrt{2k_b T_{b,air}/(\pi m)}$ |
| $\alpha_b, \alpha_t$ | bottom (top) surface thermal accommodation coefficient, assumed to be 0.7 for all calculations |
| $f$ | hole filling fraction (fraction of the total horizontal area of the structure that is holes) |
| $w$ | wall filling fraction (fraction of the total internal volume of the structure that is wall material) |
| $H$ | structure height (distance between top and bottom surfaces) |
| $T$ | ambient temperature |
| $P$ | ambient pressure |
| $v$ | ambient mean gas speed |
| $d$ | thermal boundary layer thickness |
| $k_{air}$ | bulk thermal conductivity of air, 0.0284 W/(m K) |
| $k_{mat}$ | thermal conductivity of the scaffold material, 1.8 W/(m K) for ALD alumina [1] |
| $h_b, h_t,$ $h_m, h_{m,air}$ | heat transfer coefficients in air W/(m² K) from the bottom layer to ambient, the top layer to ambient, between top and bottom layers, and between surface air layers, respectively |
| $\lambda$ | mean free path |
| $S(t)$ | solar constant, 1366 W/m², weighted by a positive half-sine wave with 24 h period |
| $A$ | average terrestrial albedo, 0.3 |
| $\epsilon_{vis,b}, \epsilon_{vis,t}$ | solar band emissivity of bottom (top) layer |
| $\epsilon_{IR,b}, \epsilon_{IR,t}$ | longwave (IR) band emissivity of the bottom (top) layer of the structure |
| $T_e$ | terrestrial thermal upwelling blackbody temperature, 255 K |
| $c_P$ | heat capacity of air, 1000 J/(kg K) |
| $\sigma$ | Stefan-Boltzmann constant, $5.67 \times 10^{-8}$ W/(m² K⁴) |
| $k_B$ | Boltzmann constant, $1.38 \times 10^{-23}$ J/K |
| $R$ | ideal gas constant for dry air, 287.1 J/(kg K) |
| $m$ | mean molecular mass of air, $4.81 \times 10^{-26}$ kg |
| $g$ | gravitational acceleration on Earth, 9.8 m/s² |
| $u$ | settling velocity of a plate |
| $s$ | side length of a square plate |
| $G$ | dimensionless parameter characterizing the aspect ratio of the area of a plate normal to a flow of gas |
| $AD$ | area density of a plate |
| $\mu$ | dynamic viscosity of air |
| $\nu$ | kinematic viscosity of air |
| $D$ | diameter of disk structure |
| $\beta$ | structure's tilt angle with respect to horizontal |
| $\Delta T(x, y)$ | temperature difference between PAS layers, $T_b - T_t$, as a function of horizontal position $(x, y)$ |
| $T_w$ | temperature of superstructure walls |
| $\delta$ | thickness of PAS alumina sandwich (face sheets and post walls) |

**Uniform pressure assumption**

To check the validity of our assumption that the pressure $P$ is uniform at all regions in the models, we created a simplified version of the models that assumes the pressure between the two sides of the structure is a dependent variable and the temperatures $T_{b,air}$ and $T_{t,air}$ are independent variables. $T_b$ and $T_t$ are not relevant parameters for this model and are therefore not defined. Supplementary Figure 1 shows a vertical cross-section of the "three-chamber" structure represented by this model. We use $T_{b,air} = 251$ K and $T_{t,air} = 250$ K to account for the typical order 1 K difference between the surface air layers for an optimized structure at 25 km. At this altitude, the model returned a negligibly larger thrust (a factor of $< 10^{-4}$ increase) on the structure than if $P$ were uniform throughout the model. This indicates that the thermomolecular air pressure differences caused by a thermal gradient within the structure have a negligible effect on the thrust and can be ignored.

**Boundary layer thickness calculation**

We calculate $d$ as the thermal boundary layer thickness given by forced convection on a heated thin disk aligned horizontally, normal to a laminar airflow:

$$d = 5\sqrt{\frac{D\,v}{2\,u}}\,0.71^{-\frac{1}{3}}$$

where $D$ is the disk diameter, $v$ is the kinematic viscosity of air, $u$ is the airflow speed, and 0.71 is the Prandtl number for air [2]. As discussed in the main text, this calculation was supplanted by the assumption $d = D$ for the structures we model.

**Stokes flow regime assumption**

Could a tilted structure fly fast enough horizontally to overcome prevailing winds and maintain a roughly constant position over the ground? If so, this station-keeping structure could communicate with a single ground station. The horizontal component of the lofting force normal to a structure determines its horizontal motion. The drag is determined by the Reynolds number $Re$, which is highly dependent on small tilt angles about 0°. We calculated $Re$ for a tilted structure using $\sin(\beta) + H\cos(\beta)$ as the structure's characteristic length scale (Supplementary Fig. 2).

Microscale structures are solidly in the Stokes flow regime ($Re \lesssim 1$) for all tilt angles in the stratosphere [3]. Structures of order 10 cm wide, however, enter the transition flow regime ($1 \lesssim Re \lesssim 1000$) when their speed relative to ambient winds is greater than roughly 0.5 m/s or the tilt angle is greater than roughly 1°. While transitional flow may be common, turbulent flow ($Re \gtrsim 1000$) will be rare. Turbulent patches are highly intermittent in the stratosphere; tracer particles encounter a new stratospheric turbulent patch on average once daily [4]. These patches are of order 50 m tall to 100 m wide [5], small enough for devices with active attitude adjustment to plausibly navigate. Our calculations therefore assume Stokes flow for both horizontal and vertical motion.

**Stokes drag equations**

In the Stokes regime, horizontal drift speed is independent of small tilt angles because the horizontal components of the lofting and drag forces are both proportional to the inclination [3]. However, the Stokes drag coefficient for slightly tilted plates is not well known. We examined two formulae, both assuming Stokes drag (see next section), for the settling velocity $u$ of a thin plate. Cortes et al. [6] found that the holes within the nanocardboard structure have no significant effect on the settling velocity. They give the first formula:

$$u = \frac{s^2(g\,AD - F_p)}{\left(1 + \frac{G\,Kn(s)}{6}\right)^{-1}(G\,\mu\,s + 2\,\rho\,v\,s^2)}.$$

For this paper, we used $G = 16$, the value reported by Cortez, et al. for a flat, circular plate normal to the air flow. Since we are not considering a square plate, we assume $s^2 = \pi\,(D/2)^2$ where $D$ is the diameter of our disk structure. The above equation contains an interpolation between the free-molecular and continuum regime drag force equations and is therefore suitable for all flow regimes. The second formula for settling velocity, given by Keith [3], is

$$u = \frac{(g\,AD - F_p)v}{4\,P}.$$

For the same structure in the absence of photophoretic forces, we found these formulae are within 20% of each other at all altitudes. Both are suitable for calculating the vertical settling velocity for our structures, where a structure is aligned perpendicularly to the flow. However, there is a dearth of data available for plates aligned parallel to or at a shallow angle to the incident flow. How the Stokes drag coefficient (proportional to $G$) changes as the angle of the plate changes is not well studied. We therefore used the above formulae for horizontal motion calculations while only replacing $s \to s \sin(\beta) + H \cos(\beta)$. A more accurate measure of the horizontal velocity of plates requires CFD calculations where the Stokes drag coefficient as a function of plate attitude can be determined. To avoid singularities at $g\,AD = F_p$, we simply assumed $F_p = 0$ when calculating horizontal and vertical motion as a function of time and $d$ (see below). CFD calculations would also be able to study the temporal feedback from $F_p \neq 0$.

**Horizontal and vertical motion**

The above drag formulae show a 10 cm diameter structure's horizontal velocity increases exponentially from order 0.1 m/s at 20 km to 1 m/s at 35 km. There are typically some stratospheric regions with average total wind speeds below 1 m/s [7,8] so station keeping using horizontal motion might be feasible at some times. Alternately, structures might station-keep by adjusting float altitude to take advantage of wind shear as is done for stratospheric balloons [9,10]. To control altitude within an operational range, a structure could tilt sufficiently to reduce lofting force so that adjustment of average tilt angle would allow limited altitude control.

We calculate a structure's altitude over time by iteratively solving the balance among the lofting, gravitational, and Stokes drag forces. Our reference device with total area density 40 g/m² will levitate at 25 km indefinitely, bobbing up and down within a vertical diurnal range of 700 m.

**FEA of PAS cells**

We performed large deformation FEA on different PAS cells to determine (1) what cell size the expected forces during flight would fracture the alumina, and (2) the effect of walls on the stress and displacement of the face sheets. For each FEA run we assumed the model structure was made of 100 nm thick alumina with fracture stress of 1 GPa, Young's modulus of 170 GPa, and Poisson's ratio of 0.21 [11–13].

We first modeled a cell of a preliminary hexagonal PAS design that is 520 µm on a side with $H = 40$ µm, $w = 10^{-3}$, $f = 0.3$, and area density 1 g/m² (Supplementary Fig. 3a). Though our fabrication procedure can create holes in the face sheets without walls, it can only create cylindrical posts if they are concentric with circular face sheet holes. Supplementary Figure 3a shows the preliminary design where every face sheet hole is concentric with a post, which would be detrimental to the lofting force but is insignificant to this structural analysis (see end of this section). Applying 10 Pa ($100g$) to the top surface induces a maximum stress of 17.4 MPa, roughly 2 % of the fracture stress (Supplementary Fig. 3b). Stress generally scales inversely with the square of area density, linearly with area, and linearly with pressure in the elastic regime [14]. Therefore, the maximum expected load of 0.4 Pa on our reference PAS (with area density 0.5 g/m² and side length 5 mm) would induce roughly 1 % of the fracture stress.

Next, we modeled a single PAS face sheet as a 1 mm² square with $f = 0.4$ to represent a PAS with $w = 0$ (two membranes separated by $H$) rigidly fixed on along its perimeter. We applied a uniform pressure of 10 g/m², or 0.1 Pa, to the sheet, resulting in a maximum displacement of 0.5 µm and maximum stress of 0.5 MPa (Fig. 5a). Extrapolating to our reference structure, the same sheet under the maximum expected pressure of 0.4 Pa would have a maximum displacement of 2 µm and stress of 2 MPa. Though the expected stress is 500 times smaller than the fracture stress, the displacement is too large for our reference structure with $H = 2$ µm because the face sheets will touch. To keep sheets with separation $H = 2$ µm from touching when maximally displaced, posts would need to be placed roughly every 500 µm apart. This spacing is sparse enough to (1) insignificantly affect the temperature gradient, and thus the lofting force, due to horizontal temperature gradients in the top sheet and (2) keep $w < 10^{-4}$ with 1 µm radius holes. This is also true for posts spaced as low as 60 µm apart; we chose this value for our reference design.

Note that the maximum stress expected on our reference structure is somewhat insensitive to the presence of walls. We do not expect significant shear stresses on the PAS so long as the SS is adequately stiff. Walls provide little bending stiffness when a force is applied normal to the face of the PAS. So long as walls are spaced adequately close to each other to limit the local relative bending of the face sheets, the "no-straight-line rule," a benchmark of high bending stiffness in nanocardboard-like structures [15,16], need not be satisfied in our reference PAS.

**Lofting force generated by holes concentric with cylindrical posts**

We found $\Delta T$, and thus the lofting force, reached 90 % of their maximum roughly 10 µm from the edge of the post. The probability density function for a gas molecule's free path $x$ is $\frac{1}{\lambda}e^{-x/\lambda}\,dx$. We multiplied this formula by $\Delta T(x, 0)$, integrated from 0 to infinity, and divided by $\int_0^\infty \Delta T(x, 0)dx$ to calculate the mean temperature of a gas molecule incident on the hole. We found this mean temperature was 45 % of the maximum value of $\Delta T$. This indicates that for a given 100 nm thick post and hole with radius $> 100$ nm and $< \lambda$, the lofting force would be roughly twice as large in a structure where the post is far ($> 10$ µm) from the hole compared to concentric with the hole.

**FEA of honeycomb SS**

As discussed in the practical device design section in the main text, we performed FEA to determine the maximum size of the SS honeycomb pattern that could withstand being fixed on one side without itself buckling or fracturing the PAS under gravity.

We modeled our reference SS design; a honeycomb with a height of 1 mm, top and bottom layer thicknesses of 1.5 µm, vertical layer thickness of 300 nm, area density of 4.3 g/m², hexagon width (edge to edge) of 1 mm, and area filling of 10 % as a 1 mm wide cantilever beam fixed on one side with periodic boundary conditions perpendicular to the beam (Fig. 5b and Supplementary Fig. 4a). The critical buckling moment of the beam is reached when there is a sharp drop in its flexural rigidity. The critical buckling moment for the 1 mm wide test cantilever is 0.3 N µm (Supplementary Fig. 4c).

The same cantilever beam with width $s$ mm would have a critical buckling moment of $0.3s$ N µm. Compare this to a square cantilever beam fixed on one side with maximum bending moment $s^3\,AD\,g/2$. If this moment exceeds the critical buckling moment of the cantilever, the cantilever will buckle. Setting the critical buckling moment equal to the maximum bending moment and solving gives $s = 12$ cm. A square made of our reference SS will therefore buckle under its own weight if its sides are 12 cm or longer.

As discussed in the main text, we ignore shear stresses and post-buckling behavior in our SS analysis. Should higher shear stiffness be required, one could add semicircular "ribs" at the midpoints of the hexagonal edges, herein called our modified SS design (Supplementary Fig. 4b). We modeled four cantilever beams under pure bending moments: (1) our reference SS discussed in the main text (2) a modified SS with 100 µm radius ribs and otherwise the same dimensions as (1), (3) our reference SS but with 1 µm layer thickness all around, and (4) a modified SS with 100 µm radius ribs and otherwise the same dimensions as (3). (1) and (2) both have area densities of about 4.5 g/m$^2$ while (3) and (4) both have area densities of about 12 g/m$^2$. Supplementary Figure 4c shows how the bending stiffness of each beam changes with the applied moment. Supplementary Figure 4d shows how the apparent bending stiffness, a combined metric of both bending and shear stiffness, changes with the beam length. The apparent bending stiffness is a measure of the relative contributions of shear and bending to the total displacement of the structure. For beam lengths where the apparent bending stiffness is constant, bending is the dominant contribution to the deflection. We calculate the apparent bending stiffness using the method in Kim et al. [16]. In general, the modified SS beams have greater shear stiffness than our reference SS beams, shown by the modified SS beams' apparent bending stiffnesses reaching constant values at shorter beam lengths than our reference SS beams. However, the modified SS designs also have smaller pure bending stiffnesses.

Alumina plates with micron-scale thickness have fracture stresses around 200-300 MPa [17]. We found the maximum stress on the reference SS beam under its critical buckling moment is 5 MPa. Since the fracture stress is much greater than the critical buckling stress, we focus on buckling as the primary failure mode.

The angle of curvature of the beam under its critical buckling moment is $5.7 \times 10^{-4}$ radians. We applied this curvature to the PAS sheet of Fig. 5a. The curvature induced a displacement of roughly 600 nm and maximum stress of roughly 0.6 MPa within the PAS; over $10^3$ times less than the stress needed to fracture it.

**Proposed method of fabricating the PAS and SS simultaneously**

The PAS and honeycomb SS can be fabricated simultaneously using the following procedure (Supplementary Fig. 5). (1) We start with a silicon wafer with the same thickness as the desired superstructure height. (2) Deep reactive ion etching (DRIE) of the silicon creates the honeycomb superstructure pattern. (3) Further DRIE opens holes along the junction of the superstructure and PAS regions. (4) ALD deposits alumina of the desired superstructure thickness on all exposed areas of silicon. (5) Reactive ion etching (RIE) removes the alumina in the PAS regions where no posts are desired but leaves alumina in the superstructure and junction regions. DRIE then removes the exposed silicon. (6) The desired absorption layer (e.g. chromium and alumina) is deposited by alternating sputtering and ALD. Ozone plasma directionally cleans any holes that were covered by the coating. The desired radiatively cooling layer is deposited on the top layer, likely using chemical vapor deposition (CVD) depending on the material. (7) XeF$_2$ gas etches the remaining silicon, leaving our completed structure.

**SDS trusses**

SkyCiv structural engineering software was used to find the response of three-legged SDS trusses (Fig. 5C) to the weight of the entire practical structure, roughly 300 mg. The 10 cm truss weighed 10 mg and was made of I-beams with 0.25 mm depth and width, and 10 µm flange width, web width, and fillet radius. We also modeled a 20 cm truss weighing 230 mg that used I-beams with 0.4 mm depth and width, and 16 µm flange width, web width, and fillet radius. The I-beams are assumed to be made of carbon fiber reinforced plastic, with Young's modulus 150 GPa, Poisson's ratio 0.2, and density 3500 kg/m$^3$. The trusses are pinned on 3D-hinges at the bases of their legs. The load was applied as a point force on the top of the trusses. Both trusses returned a safety factor of about 40, meaning buckling would have occurred if the load were multiplied by that amount.

The SDS will encounter two major loads throughout the device's use. The first is $2g$ on the entire structure, or roughly 600 mg on the 10 cm diameter structure, during transport. The second is $60g$ on the PAS and SS, or $1.2g$ on the entire structure, during deployment. Forces encountered during flight will only heavily affect the PAS and SS because the SDS is not fixed in space. Our analysis shows the SDS trusses above have safety factors of 10 when they encounter their largest load of $2g$.

**Deployment and exposure in the stratosphere**

Turbulent forces make deployment from aircraft implausible. Deployment might be achieved using weather balloon technology that can easily lift 1 kg to the stratosphere. Devices could be fabricated and transported while attached to a carrying frame with piezoelectrically controlled latches or fusible wax joints.

Lofted structures must endure constant exposure to UV radiation, sulfate aerosols, and ozone. Inert structural components like the alumina sandwich and SS will be largely unaffected. However, possible surface coatings and DLW materials such as organic polymers and metals must be treated to prevent degradation. Sulfate aerosol condensation may accelerate polymer degradation, damage electrical components, and block holes in the PAS over time.

**Supplementary References**

1. Cappella, A. *et al.* High Temperature Thermal Conductivity of Amorphous Al$_2$O$_3$ Thin Films Grown by Low Temperature ALD. *Adv. Eng. Mater.* **15**, 1046–1050 (2013).
2. Schlichting, H. *Boundary-Layer Theory*. (Springer, 1979).
3. Keith, D. W. Photophoretic levitation of engineered aerosols for geoengineering. *Proc. Natl. Acad. Sci. U.S.A.* **107**, 16428–16431 (2010).
4. Vanneste, J. Small-scale mixing, large-scale advection, and stratospheric tracer distributions. *J. Atmos. Sci.* **61**, 2749–2761 (2004).
5. Haack, A., Gerding, M. & Lübken, F.-J. Characteristics of stratospheric turbulent layers measured by LITOS and their relation to the Richardson number. *J. Geophys. Res. Atmos.* **119**, 10,605-10,618 (2014).
6. Cortes, J. *et al.*, Photophoretic levitation of macroscopic nanocardboard plates. *Adv. Mater.* **32**, 1906878 (2020).
7. Brady, E. A. Zonal mean zonal wind - stratospheric perspective. *ECMWF* https://sites.ecmwf.int/era/40-atlas/docs/section_D25/parameter_zmzwsp.html (2006).

8. Flury, T., Wu, D. L. & Read, W. G. Variability in the speed of the Brewer–Dobson circulation as observed by Aura/MLS. *Atmos. Chem. Phys.* **13**, 4563–4575 (2013).
9. Bellemare, M. G. *et al.* Autonomous navigation of stratospheric balloons using reinforcement learning. *Nature* **588**, 77–82 (2020).
10. Modica, G. D., Nehrkorn, T. & Myers, T. An investigation of stratospheric winds in support of the high altitude airship. *13th Conference on Aviation, Range and Aerospace Metrology* (AER, 2007).
11. Ruoho, M. *et al.* Thin-film engineering of mechanical fragmentation properties of atomic-layer-deposited metal oxides. *Nanomaterials* **10**, 558 (2020).
12. Rontu, V. *et al.* Elastic and fracture properties of free-standing amorphous ALD $Al_2O_3$ thin films measured with bulge test. *Mater. Res. Express* **5**, 046411 (2018).
13. Berdova, M. *et al.* Fracture properties of atomic layer deposited aluminum oxide free-standing membranes. *J. Vac. Sci. Technol. A* **33**, 01A106 (2015).
14. Roark, R. J., Young, W. C. & Budynas, R. G. *Roark's formulas for stress and strain*. (McGraw-Hill, 2002).
15. Lin, C. *et al.* Nanocardboard as a nanoscale analog of hollow sandwich plates. *Nat. Commun.* **9**, 4442 (2018).
16. Kim, J. *et al.* Ultralight and ultra-stiff nano-cardboard panels: Mechanical analysis, characterization, and design principles. Preprint at https://arxiv.org/abs/2211.02803 (2022).
17. Orlovská, M., Chlup, Z., Bača, Ľ., Janek, M. & Kitzmantel, M. Fracture and mechanical properties of lightweight alumina ceramics prepared by fused filament fabrication. *J. Eur. Ceram. Soc.* **40**, 4837–4843 (2020).
30

**Supplementary Figures**

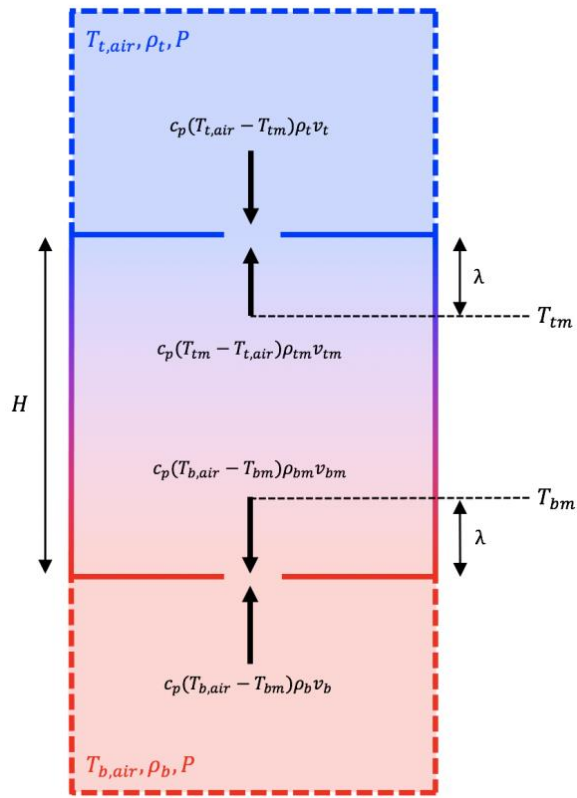

**Supplementary Figure 1.** "Three-chamber" structure where the temperatures of the top and bottom surfaces are in thermal equilibrium with their surroundings and the middle chamber's pressure is the dependent variable of interest. The heat flow terms are shown with arrows representing their directionality.

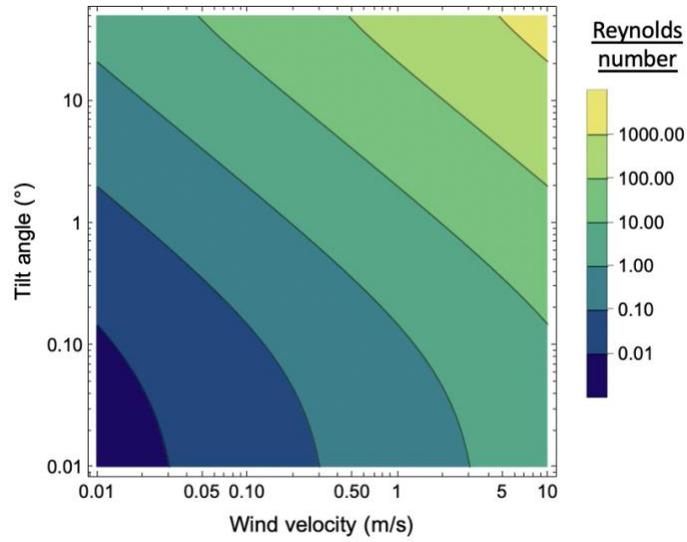

**Supplementary Figure 2.** Reynolds number for a 10 cm diameter disk at 25 km as a function of tilt angle $\beta$ and wind velocity relative to the disk.

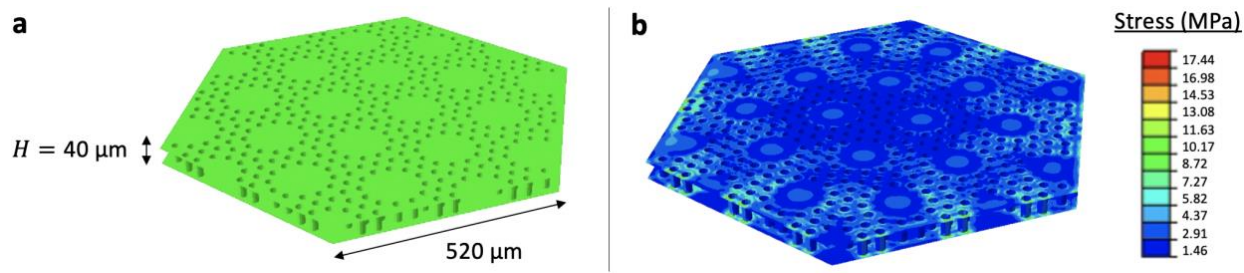

**Supplementary Figure 3.** FEA of a preliminary PAS cell design. **a** Input model of the design with all holes concentric with 100 nm-thick posts for FEA. **b** Stress map of the structure when fixed on all sides and subject to 10 Pa on the top surface.

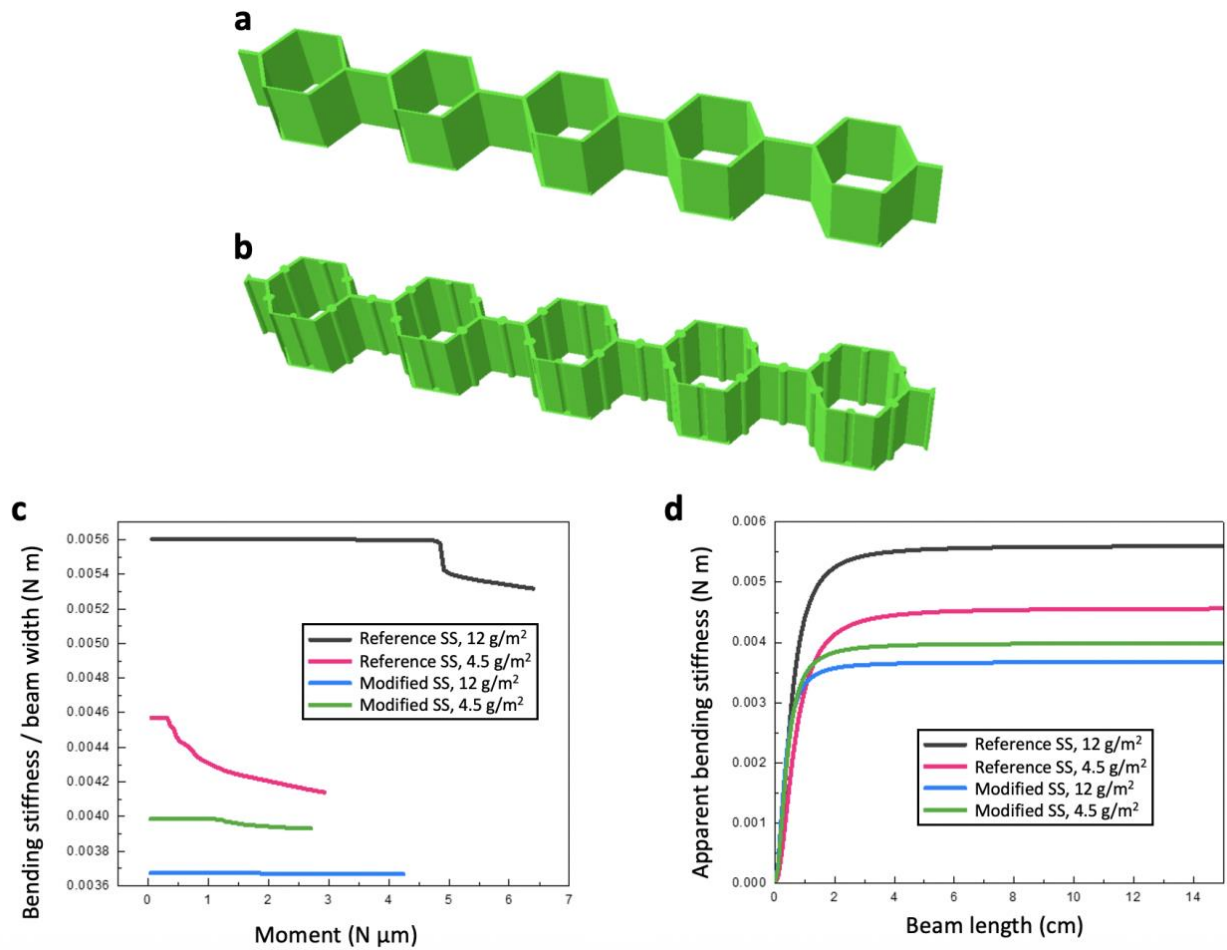

**Supplementary Figure 4.** FEA of four SS designs. **a** Reference SS: a honeycomb design made with beams that have a rectangular vertical cross section. **b** Modified SS: a honeycomb design that has semicircular ribs at the midpoint of the sides of the hexagons. **c** Plots of bending stiffness divided by the width of the beam as a function of moment applied at the end of the beam. The four designs are: (pink) our reference SS discussed in the main text, (green) a modified SS with 100 µm radius ribs, (black) our reference SS but with 1 µm layer thickness all around, and (blue) a modified SS with 100 µm radius ribs and 1 µm layer thickness all around. The designs with 1 µm layer thickness have area densities of about 12 g/m$^2$ while the others have area densities of about 12 g/m$^2$. The critical buckling moment of each design is the moment at which the bending stiffness drops suddenly. **d** Plots of apparent bending stiffness as a function of beam length; see Kim et al. [16]. For beam lengths where the apparent bending stiffness is constant, bending (rather than shear) is the dominant contribution to the deflection.

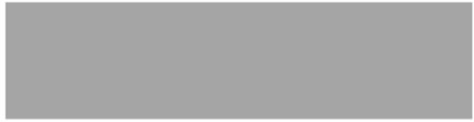
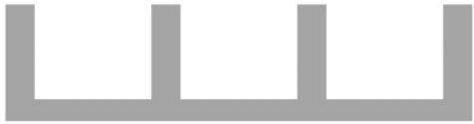
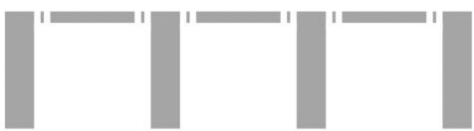
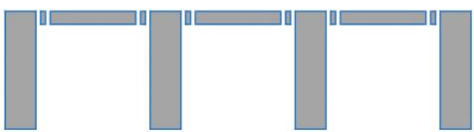
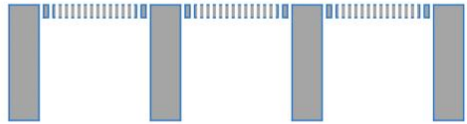
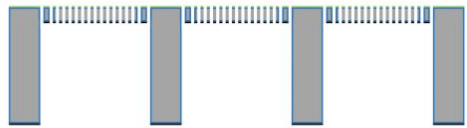
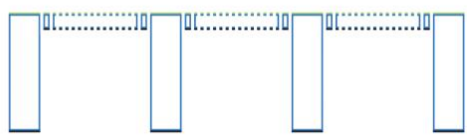

**Supplementary Figure 5.** Procedure for fabricating PAS cells and a hexagonal superstructure concurrently. The materials used are silicon (gray), alumina (blue), a radiatively warming coating (black), and a radiatively cooling coating (green).

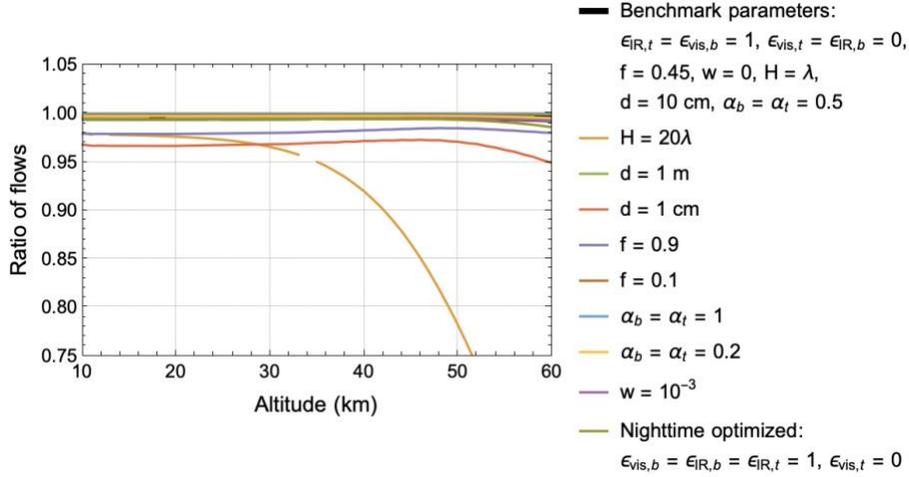

**Supplementary Figure 6.** The ratio of net flow of mass through the bottom layer to net flow through the top layer,

$$\left(\sqrt{T_{b,air}} + \sqrt{T_{bm}}\right)\left(\frac{1}{\sqrt{T_{bm}}} - \frac{1}{\sqrt{T_{b,air}}}\right) \Big/ \left(\left(\sqrt{T_{t,air}} + \sqrt{T_{tm}}\right)\left(\frac{1}{\sqrt{T_{t,air}}} - \frac{1}{\sqrt{T_{tm}}}\right)\right),$$

in the $H \gg \lambda$ model, plotted for variations in each of the model's input parameters, all at peak daytime. Each plot assumes the benchmark parameters except for the noted variations.

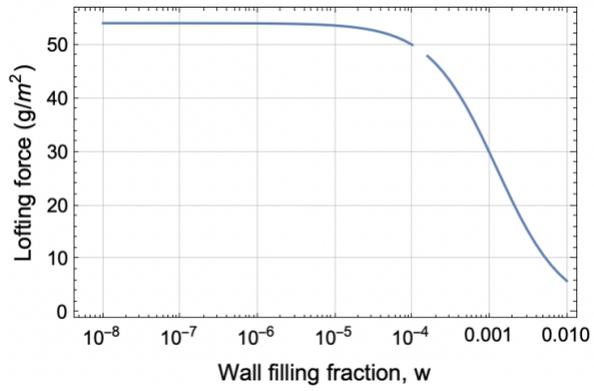

**Supplementary Figure 7.** Time-averaged lofting force vs. wall filling fraction for a structure with benchmark parameters at 25 km.

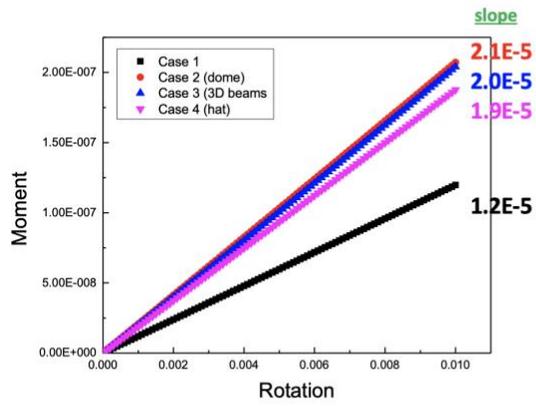 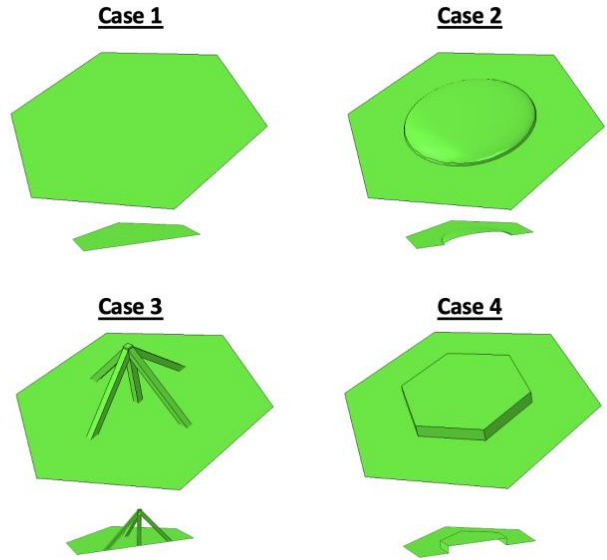

**Supplementary Figure 8.** FEA analysis on four different horizontal profiles with base sheet thickness of 4 µm. Case 2 shows a 200 µm wide, 20 µm tall half-spheroid structure. We assume the structure is made of alumina with Young's modulus of 170 GPa and Poisson's ratio of 0.2.